\documentclass{emulateapj}

\usepackage{graphics}
\usepackage{epsfig}
\usepackage{natbib}
\usepackage[hidelinks]{hyperref}
\usepackage{asmmath} 
\shorttitle{The MUSCLES Treasury Survey Overview}
\shortauthors{France et al.}
\begin{document}

\title{The MUSCLES Treasury Survey I: Motivation and Overview\altaffilmark{*}}


\author{
Kevin France\altaffilmark{1},
R. O. Parke Loyd\altaffilmark{1}, 
Allison Youngblood\altaffilmark{1},
Alexander Brown\altaffilmark{2},
P. Christian Schneider\altaffilmark{3}, 
Suzanne L. Hawley\altaffilmark{4},
Cynthia S. Froning\altaffilmark{5}, 
Jeffrey L. Linsky\altaffilmark{6},  
Aki Roberge\altaffilmark{7}, 
Andrea P. Buccino\altaffilmark{8}
James R. A. Davenport\altaffilmark{9,10},
Juan M. Fontenla\altaffilmark{11},
Lisa Kaltenegger\altaffilmark{12}
Adam F. Kowalski\altaffilmark{13},
Pablo J. D. Mauas\altaffilmark{8}
Yamila Miguel\altaffilmark{14},
Seth Redfield\altaffilmark{15},
Sarah Rugheimer\altaffilmark{16}
Feng Tian\altaffilmark{17},
Mariela C. Vieytes\altaffilmark{18}
Lucianne M. Walkowicz\altaffilmark{19}
Kolby L. Weisenburger\altaffilmark{4} 
}

\altaffiltext{*}{Based on observations made with the NASA/ESA $Hubble$~$Space$~$Telescope$, obtained from the data archive at the Space Telescope Science Institute. STScI is operated by the Association of Universities for Research in Astronomy, Inc. under NASA contract NAS 5-26555.}


\received{December 18, 2015}
\revised{}
\accepted{February 20, 2016}

\begin{abstract}
Ground- and space-based planet searches employing radial velocity techniques and transit photometry have detected thousands of planet-hosting stars in the Milky Way.  With so many planets now discovered, the next step towards identifying potentially habitable planets is atmospheric characterization.  While the Sun--Earth system provides a good framework for understanding atmospheric chemistry of Earth-like planets around solar-type stars, the observational and theoretical constraints on the atmospheres of rocky planets in the habitable zones around low-mass stars (K and M dwarfs) are relatively few.  The chemistry of these atmospheres is controlled by the shape and absolute flux of the stellar spectral energy distribution, however, flux distributions of relatively inactive low-mass stars are poorly known at present.  To address this issue, we have executed a panchromatic (X-ray to mid-IR) study of the spectral energy distributions of 11 nearby planet hosting stars, the {\it Measurements of the Ultraviolet Spectral Characteristics of Low-mass Exoplanetary Systems} (MUSCLES) Treasury Survey.   The MUSCLES program consists of contemporaneous observations in the X-rays with $Chandra$ and {\it XMM-Newton}, ultraviolet observations with $Hubble$, and visible observations from $Hubble$ and ground-based observatories.  Infrared and astrophysically inaccessible wavelengths (EUV and Ly$\alpha$) are reconstructed using stellar model spectra to fill-in gaps in the observational data.  

In this overview and companion papers describing the MUSCLES survey, we show that energetic radiation (X-ray and ultraviolet) is present from magnetically active stellar atmospheres at all times for stars as late as M5.  Emission line luminosities of \ion{C}{4} and \ion{Mg}{2} are strongly correlated with band-integrated luminosities and we present empirical relations that can be used to estimate broadband FUV and XUV ($\equiv$~X-ray + EUV) fluxes from individual stellar emission line measurements.  We find that while the slope of the spectral energy distribution, FUV/NUV, increases by approximately two orders of magnitude form early K to late M dwarfs ($\approx$~0.01~to~1), the absolute FUV and XUV flux levels at their corresponding habitable zone distances are constant to within factors of a few, spanning the range 10~--~70 erg cm$^{-2}$ s$^{-1}$ in the habitable zone.  Despite the lack of strong stellar activity indicators in their optical spectra, several of the M dwarfs in our sample show spectacular flare emission in their UV light curves.  We present an example with flare/quiescent ultraviolet flux ratios of order 100:1, where the transition region energy output during the flare is comparable to the total quiescent luminosity of the star $E_{flare}$(UV)~$\sim$~0.3 $L_{*}$$\Delta$$t$ ($\Delta$$t$ = 1 second).  Finally, we interpret enhanced $L(line)$/$L_{Bol}$ ratios for \ion{C}{4} and \ion{N}{5} as tentative observational evidence for the interaction of planets with large planetary mass-to-orbital distance ratios ($M_{plan}$/$a_{plan}$) with the transition regions of their host stars.   

\end{abstract}

\keywords{planetary systems --- stars: individual (GJ 1214,  GJ 876, GJ 581, GJ 436, GJ 176, GJ 667C, GJ 832, HD 85512, HD 40307, $\epsilon$~Eri, HD 97658) --- stars: activity --- stars: low-mass}


\section{Introduction}

The $Kepler$ mission and ground-based planet searches have detected thousands of exoplanets within the Milky Way and have demonstrated that approximately one-in-six main sequence FGK stars hosts an Earth-size planet (with periods up to 85 days; Fressin et al. 2013).\nocite{fressin13}
One of the highest priorities for astronomy in the coming decades is the characterization of the atmospheres, and possibly the surfaces, of Earth-size planets in the Habitable Zones (HZs, where liquid water may exist on terrestrial planet surfaces) around nearby stars.  An intermediate step towards the discovery of life on these worlds is the measurement of atmospheric gases that may indicate the presence of biological activity.  These gases are often referred to as biomarkers or biosignatures.   

However, the planetary effective surface temperature alone is insufficient to accurately interpret biosignature gases when they are observed in the coming decades.  The dominant energy input and chemistry driver for these atmospheres is the stellar spectral energy distribution (SED).  The ultraviolet (UV) stellar spectrum which drives and regulates the upper atmospheric heating and chemistry on Earth-like planets, is critical to the definition and interpretation of biosignature gases (e.g., Seager et al. 2013), and may even produce false-positives in our search for biologic activity (Hu et al. 2012; Tian et al. 2014; Domagal-Goldman et al. 2014).  

The nearest potentially habitable planet is likely around an M dwarf at $d$~$<$~3 pc \citep{dressing15}, the nearest known Earth-size planet orbits an M dwarf (GJ 1132b, $R_{p}$~=~1.2 $R_{\oplus}$, $d$~=~12 pc; Berta-Thompson et al. 2015), and the nearest known Super-Earth mass planets in habitable zones orbit M and K dwarfs, making planetary systems around low-mass stars prime targets for spectroscopic biomarker searches (see also Cowan et al. 2015).\nocite{cowan15,berta15}   
The low ratio of stellar-to-planetary mass more readily permits detection of lower mass planets using the primary detection techniques (radial velocity and transits).  Moreover, the HZ around a star moves inward with decreasing stellar luminosity.  These factors make potentially habitable planets easier to detect around M and K dwarfs.   The importance of M dwarf exoplanetary systems is underscored by recent $Kepler$ results and radial velocity measurements showing that between $\sim$~10~--~50\% of M dwarfs host Earth-size planets (0.5~--~1.4~$R_{\oplus}$) in their HZs (Dressing \& Charbonneau 2015; Kopparapu 2013; Bonfils et al. 2013).  Furthermore, approximately 70\% of the stars in the Milky Way are M dwarfs, so rocky planets around low-mass stars likely dominate the planet distribution of the Galaxy.  Theoretical work has shown that planets around M dwarfs could be habitable despite their phase-locked orbits (Joshi 2003) and dynamic modeling of transiting systems reveals that most systems permit stable orbits of Earth-mass planets in the HZ long enough for the development of life, i.e. $\gtrsim$~1.7 Gyr (Jones \& Sleep 2010).\nocite{dressing15,kopparapu13,bonfils13}

M and K dwarfs show significantly larger temporal variability and fraction of their bolometric luminosity at UV wavelengths than solar-type stars~\citep{france13}, yet their actual spectral and temporal behavior is not well studied except for a few young ($<$ 1 Gyr), active flare stars.  The paucity of UV spectra of low-mass stars and our current inability to accurately model the UV spectrum of a particular M or K dwarf without a direct observation limits our ability to reliably predict possible atmospheric biomarkers.  Without the stellar UV spectrum, we cannot produce realistic synthetic spectra of Earth-like planets in these systems, a necessary step for interpreting biomarker gases and their potential to diagnose habitability.  Therefore, our quest to observe and characterize biosignatures on rocky planets must consider the star-planet system as a whole, including the interaction between the stellar irradiance and the exoplanetary atmosphere.

\subsection{High-energy Spectra as Photochemical Atmospheric Model Inputs}

{\it FUV and NUV Irradiance: Photochemistry and Biosignatures}~--~Spectral observations of O$_{2}$, O$_{3}$, CH$_{4}$, and CO$_{2}$, are expected to be the most important signatures of biological activity on planets with Earth-like atmospheres (Des Marais et al. 2002; Kaltenegger et al. 2007; Seager et al. 2009).  The chemistry of these molecules in the atmosphere of an Earth-like planet depends sensitively on the strength and shape of the UV spectrum of the host star~\citep{segura05}.  H$_{2}$O, CH$_{4}$, and CO$_{2}$ are sensitive to FUV radiation (912~--1700~\AA), in particular the bright HI Ly$\alpha$ line, while the atmospheric oxygen chemistry is driven by a combination of FUV and NUV (1700~--~3200~\AA) radiation (Figure 1).    

The photolysis (photodissociation) of CO$_{2}$ and H$_{2}$O by Ly$\alpha$ and other bright stellar chromospheric and transition region emission lines (e.g., \ion{C}{2} $\lambda$1335~\AA\ and \ion{C}{4} $\lambda$1550~\AA) can produce a buildup of O$_{2}$ on planets illuminated by strong FUV radiation fields.  Once a substantial O$_{2}$ atmosphere is present, O$_{3}$ is primarily created through a multi-step reaction whereby O$_{2}$ dissociation (by 1700 ~--~2400~\AA\ photons) is followed by the reaction O + O$_{2}$ + $\xi$~$\rightarrow$~O$_{3}$ + $\xi$, where $\xi$ is a reaction partner required to balance energy conservation.  O$_{3}$ photolysis is then driven by NUV and blue optical photons. Therefore, on planets orbiting stars with strong FUV and weak NUV flux, a substantial O$_{3}$ atmosphere may arise via photochemical processes alone ~\citep{segura10,hu12,gao15,sonny15}.

This strong photochemical source of O$_{3}$ may be detectable by future space observatories designed to carry out direct atmospheric spectroscopy of rocky planets (e.g., the HDST or LUVOIR mission concepts), and may be misinterpreted as evidence for biologic activity on these worlds.  Therefore, characterization of the stellar FUV/NUV ratio is an essential complement to spectroscopy of exoplanet atmospheres to control for potential false-positive ``biomarkers''.   Furthermore, it has been shown that the abundances of water and ozone, as well as the atmospheric equilibrium temperature, can respond to changes in the stellar flux on timescales ranging from minutes to years~\citep{segura10}.    A detailed knowledge of the absolute flux level and temporal behavior stellar spectrum is important for understanding the evolution of potentially habitable atmospheres.




{\it X-ray and EUV Irradiance: Atmospheric Heating and Mass-Loss}~--~The ratios of X-ray to total luminosity of M dwarfs are orders of magnitude higher ($\gtrsim$~10~--~100~$\times$) than those of the present day Sun (Poppenhaeger et al. 2010), and the smaller semi-major axes of the HZ around M dwarfs means that X-ray effects on HZ planets will likely be more important than on HZ planets orbiting solar-type stars (Cecchi-Pestellini et al. 2009).   Soft X-ray heating of planetary atmospheres enhances evaporation and atmospheric escape (Scalo et al. 2007; Owen \& Jackson 2012) which may impact the long-term stability of an exoplanetary atmosphere. 
Recent works suggest that the influences of early evolution of low mass stars and XUV heating could lead to a bi-modal distribution of water fractions on Earth-mass planets in the HZ of M dwarfs~\citep{tian15}.   In order to model the atmosphere as a system, we require inputs for both heating (soft X-ray and EUV, see below) and photochemistry (FUV and NUV).  

Extreme-UV (EUV; 100~$\lesssim$~$\lambda$ $\lesssim$~911~\AA) photons from the central star are an important source of atmospheric heating and ionization on all types of extrasolar planets.  For terrestrial atmospheres, increasing the EUV flux to levels estimated for the young Sun ($\approx$~1 Gyr; Ayres 1997) can increase the temperature of the thermosphere by a factor of~$\gtrsim$~10 (Tian et al. 2008), potentially causing significant and rapid atmospheric mass-loss.  Ionization by EUV photons and the subsequent loss of atmospheric ions to stellar wind pick-up can also drive extensive atmospheric mass-loss on geologic time scales (e.g., Rahmati et al. 2014 and references therein).\nocite{rahmati14}   Estimates of the incident EUV flux are therefore important for evaluating the long-term stability of a HZ atmosphere; however, a direct measurement of the EUV irradiance from an exoplanetary host star is essentially impossible because    interstellar hydrogen removes almost all of the stellar EUV flux for virtually all stars except the Sun.  
The stellar EUV energy budget contains contributions from both the transition region (Lyman continuum as well as helium and metal line emission in the 228~--~911~\AA\ bandpass) and the corona.  FUV emission lines (Ly$\alpha$, \ion{C}{4}, and \ion{Si}{4}) are required to estimate the contribution of the transition region to the EUV flux (Fontenla et al. 2011; Linsky et al. 2014), while X-ray data are necessary to constrain the contribution of the corona (e.g., Sanz-Forcada et al. 2011).  

\begin{figure}
\begin{center}
\epsfig{figure=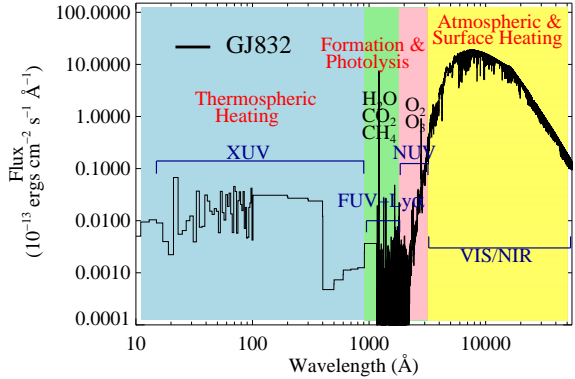,angle=90,width=3.0in}
\vspace{+0.0in}
\caption{
Panchromatic spectrum of GJ 832, illustrating the influence of each spectral bandpass on an Earth-like planet orbiting this star.  GJ 832 has a super-Earth mass planet located in the HZ~\citep{wittenmyer14}.  }
\end{center}
\end{figure}

\subsection{Variability on Timescales of Minutes-to-Hours: Atmospheric Abundances}

An important measurement relating to the habitability of extrasolar planets is the time variability of the energetic incident radiation.  
While most of the quiescent UV emission from M dwarfs comes from emission lines, continuum emission can become the dominant UV luminosity source during flares (Kowalski et al. 2010).  The relative UV emission line strengths also vary during flares (e.g., Hawley et al. 2003; Osten et al. 2005; Loyd \& France 2014).  
Thus, molecular species in the atmospheres of HZ planets will be ``selectively pumped'' during quiescent periods; only species that have spectral coincidences with stellar emission lines will be subject to large energy input from the host star.  However, during flares with strong continuum emission, the relative excitation and dissociation rates relative to quiescent periods could change radically. 
Therefore, temporally and spectrally resolved observations are essential for understanding the impact of time variability on HZ planetary atmospheres.   The amplitude and frequency of flare activity on older M-star exoplanetary hosts is completely unexplored, although GALEX NUV imaging observations suggests that flares may significantly alter the steady-state chemistry in the atmospheres of planets in the HZ (Welsh et al. 2006).  

Impulsive UV events are also signposts for energetic flares associated with large ejections of energetic particles.  Segura et al. (2010) have shown that energetic particle deposition into the atmosphere of an Earth-like planet without a magnetic field during a large M dwarf flare can lead to significant atmospheric O$_{3}$ depletions ($>$ 90\% for the most extreme flares).  \citet{buccino07} also studied the impact of a series of lower intensity flares from highly active stars.  These events could alter the atmospheric chemistry and increase the penetration depth of UV photons that are damaging to surface life.   The impact of a single flare may be detrimental to the development and maintenance of life, but the potentially far more significant impact of persistent flare events has not been studied because the temporal behavior of UV flares is unexplored outside of a few extreme M dwarf flare stars (e.g., AD Leo, EV Lac, and AU Mic; e.g., Hawley et al. 1991; Osten et al. 2005; Robinson et al. 2001).\nocite{hawley91,osten05,robinson01}

\subsection{The MUSCLES Treasury Survey: An Energetic Radiation Survey of Exoplanetary Hosts}

With the previously described motivation in mind, the question that arises is ``what are the shapes and absolute flux levels of the UV stellar SEDs incident on these planetary systems?''  At present, GJ 832 is the only M dwarf for which a semi-empirical atmosphere model has been built and vetted by direct comparison with spectroscopic observations at optical, NUV, FUV, and X-ray wavelengths~\citep{fontenla15}.  There are no other stellar atmosphere models for M dwarfs that treat the chromosphere, transition region, and corona in a self-consistent manner, and none that can produce synthetic spectra for the important X-ray and ultraviolet bandpasses (5 - 3000~\AA).  

Several approaches have been taken in the literature, including assuming that the star has no UV emission (essentially, no magnetically active upper atmosphere; Segura et al. 2005; Kaltenegger et al. 2011), assuming the extreme flare environment of a star like AD Leo~\citep{segura10,wordsworth10}, or using low-S/N observations of the few flaring M dwarfs that could be observed by $IUE$~\citep{segura05,buccino07}.\nocite{kaltenegger11}  Previously available archival data are insufficient for an accurate quantitative analysis of the radiation fields incident on potentially habitable planets orbiting M dwarfs.  Low-sensitivity and contamination by geocoronal Ly$\alpha$ emission make $IUE$ observations insufficient for this work (see example in \S 2).  The lack of observational constraints from UV spectra of M dwarfs will have a major impact on how we judge whether the planets in these systems are actually inhabited.  While the need for panchromatic data has been realized for Sun-like stars (e.g., Sun in Time; Ribas et al. 2005), M dwarfs have received less attention (see also  Guinan \& Engle 2009), despite the fact that these systems dominate the planet statistics of the Galaxy.

To address the above question, we have carried out the first panchromatic survey of M and K dwarf exoplanet host stars in the solar neighborhood ($d$~$\lesssim$~20 pc).  We refer to this program as the {\it Measurements of the Ultraviolet Spectral Characteristics of Low-mass Exoplanetary Systems} (MUSCLES) Treasury Survey, a coordinated X-ray to NIR observational effort to provide the exoplanet community with empirically-derived panchromatic irradiance spectra for the study of all types of exoplanets orbiting these stars.  MUSCLES is largely based on a {\it Hubble Space Telescope} Cycle 22 treasury program and makes use of smaller guest observing programs on $HST$, {\it XMM-Newton}, $Chandra$, and several ground-based observatories.   

Our survey provides a database of the chromospheric, transition region, and coronal properties of low-mass stars hosting exoplanets, providing high-quality input for models of both Jovian and Earth-like planets as vast numbers of these systems are discovered and characterized in the next decade with missions such as $WFIRST$, $Plato$, $TESS$, and $JWST$.   While the long-term evolution of the broadband UV luminosity function of M dwarfs can be constrained with large photometric samples from $GALEX$ (e.g., Shkolnik \& Barman 2014), a further uncertainty in the temporal behavior of low-mass exoplanet host stars is the variability on time scales of years (the stellar equivalent of the solar cycle).\nocite{shkolnik14}  Given the limitations on space observatories, particularly as it is likely that we will be in the post-$HST$ and post-$Chandra$ era within the next 5~--~10 years, it is critically important to identify a set of visible-wavelength tracers (e.g., Gomes da Silva et al. 2011) that can be used to quantify the longer-term (years-to-decades) UV variability of these stars.\nocite{silva11}  The MUSCLES dataset will enable us to derive empirical relations between optical, UV, and X-ray fluxes (e.g., FUV luminosity vs. \ion{Ca}{2} and H-$\alpha$ profiles), as well as their relative behavior during flares, supporting long-term ground-based programs to study the time evolution of the energetic radiation environment.  

This paper provides an overview of the motivation for and the design of the MUSCLES Treasury Survey, as well as some initial quantitative results.  A detailed example of the need for $HST$ to carry out this work is given is Section 2.  Section 3 describes the MUSCLES target list and the description of the observing modes used in the program.   Section 4 presents the evolution of the broad-band SED with stellar effective temperature and habitable zone distance, provides scaling relations to estimate the broadband luminosity from individual spectral line measurements, and presents a first look at the intense high-energy flare behavior of these otherwise inactive stars.  Section 5 explores the interaction of the planets and host stars in these systems, and compares the UV flux measurements with predictions from coronal models.  Section 6 presents a summary of the important results from this paper.  

Details on all aspects of the data analysis and scientific results from the program can be found in the companion papers from the MUSCLES team.  \citet{youngblood15} present an analysis of the essential reconstruction of the intrinsic Ly$\alpha$ stellar emission lines and the calculations of the EUV irradiance (``Paper II''). \citet{loyd15}, ``Paper III'', describe the creation of the panchromatic spectral energy distributions (SEDs), quantification of the SEDs for exoplanet atmospheric modeling, and a description of how to download the data in machine-readable format from the Milkulski Archive for Space Telescopes\footnote{ \tt https://archive.stsci.edu/prepds/muscles/}.    \citet{fontenla15} present semi-empirical modeling of M dwarf atmospheres based in part on MUSCLES observations.  Detailed descriptions of the flare properties of the MUSCLES sources and a comparison of the UV and X-ray emission from these stars with contemporaneous ground-based photometry and spectroscopy are presented in papers by Loyd (2016~--~in prep.) and Youngblood (2016~--~in prep.), respectively, and Linsky et al. (2016) describe the kinematics of the stellar atmospheres derived from the UV emission lines in the $HST$ data.  \nocite{loyd15,loyd16,youngblood15,youngblood16,linsky16}

\section{Motivation: The Importance of High-Sensitivity, Spectrally Resolved Data for Atmospheric Modeling}

As mentioned above, the observational and theoretical understanding of the upper atmospheres of ``weakly active'' M dwarfs~\citep{walkowicz09} is insufficient for a deterministic prediction of the flux in the HZ around an M dwarf host star.  $GALEX$ data provide a statistical picture of the evolution of the UV luminosity of M dwarfs~\citep{shkolnik14}, but do not include spectral information on the specific intensity (important for accurate photoexcitation and photolysis rates of key atmospheric species), coverage of the brightest FUV and NUV spectral lines (Ly$\alpha$ and \ion{Mg}{2}, respectively), information about the flare or stellar cycle state of the star, or the connection to the X-ray and optical fluxes of the stars. \citet{shkolnik14b} provided scaling relations to calculate the Ly$\alpha$ and \ion{Mg}{2} fluxes from $GALEX$ broadband fluxes when $HST$ spectroscopy is not available. 



\begin{figure}
\begin{center}
\epsfig{figure=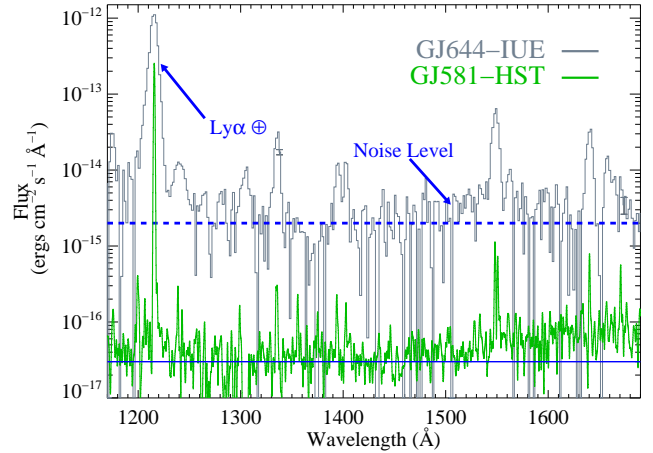,width=2.6in,angle=90}
\vspace{+0.1in}
\caption{
\label{cosovly} A comparison of the $IUE$ spectrum of the M3V star GJ 644 and the weakly active M3V exoplanet host star GJ 581 (France et al. 2013).   The $IUE$ spectrum is severely contaminated by geocoronal Ly$\alpha$ and \ion{O}{1} $\lambda$1304 emission and lacks the spectral resolution needed to permit an intrinsic profile reconstruction (see, e.g., Wood et al. 2005).  The blue dashed line represents the instrumental background level in $IUE$ FUV observations ($F_{\lambda}$~$\approx$~2~$\times$~10$^{-15}$ erg cm$^{-2}$ s$^{-1}$ \AA$^{-1}$).  Given the geocoronal emission and instrumental background levels, we estimate that approximately 80\% of the total 1160~--~1690~\AA\ FUV flux in the GJ 644 spectrum is contributed by non-stellar sources.   Higher sensitivity, higher spectral resolution observations are essential for accurately measuring the energetic radiation environment in the HZ of low-mass stars.  
 }
\end{center}
\end{figure}

To demonstrate the importance of using high-quality UV observations from $HST$, we compare in Figure 2 the observed UV flux of GJ 581 with another $\sim$~M3 (spectral types for GJ 581 range from M2.5~--~M5 in the literature) stellar spectrum taken from the $IUE$ archive (the M3V binary star GJ 644, scaled to the distance of GJ 581).  One might consider using GJ 644 as a proxy for the GJ 581 stellar radiation field before the MUSCLES data were acquired.  No weakly active M dwarfs were bright enough to be observed by $IUE$ outside of flares.   

Two discrepancies between the spectra shown in Figure 2 are immediately apparent: 1) the ``continuum flux'' level of GJ 644 is approximately two orders of magnitude larger than GJ 581.  The continuum level of the GJ 644 spectrum is approximately 2~$\times$~10$^{-15}$ erg cm$^{-2}$ s$^{-1}$ \AA$^{-1}$, which is consistent with the instrumental background equivalent flux for the $IUE$ FUV low-resolution channel.   The $HST$-COS background level is approximately 3~$\times$~10$^{-17}$ erg cm$^{-2}$ s$^{-1}$ \AA$^{-1}$, and no stellar FUV continuum is detected above this level in GJ 581.  In Paper III, we show that approximately half of the M dwarfs in the MUSCLES sample have FUV continuum detections at this flux level.  Therefore, the $IUE$ spectrum includes a large amount of instrumental noise relative to the true flux upper limits in the FUV continuum.  2) the GJ 644 Ly$\alpha$ emission line is much brighter and broader than the reconstructed Ly$\alpha$ emission from GJ 581.  The Ly$\alpha$ emission line in the $IUE$ spectrum is almost entirely geocoronal airglow emission.  The large 10$^{''}$~$\times$~20$^{''}$ oval $IUE$ aperture admits 5000 times more geocoronal emission than STIS E140M observations using the 0.2$^{''}$~$\times$~0.2$^{''}$ aperture, and the $IUE$ spectral resolution is too low to separate the interstellar absorption component from the stellar and geocoronal emission components.  Therefore, $IUE$ spectra cannot be used to compute the Ly$\alpha$ irradiance from faint, low-mass stars.  

A comparison of the instrumental background and geocoronal airglow emission with the total 1160 - 1690~\AA\ FUV flux from the $IUE$ spectrum of GJ 644 indicates that $\approx$~80~\% of the recorded counts come from non-stellar sources.  This dramatic overestimation of the FUV flux in $IUE$ M dwarf data would lead one to infer FUV/NUV flux ratios $\gtrsim$~40 when using the GJ 644 spectrum.  This is much larger than the FUV/NUV flux ratios of $\sim$~0.5~--~1 that are found for M dwarf exoplanet host stars using the higher sensitivity, higher resolution $HST$ data (Section 4.1).  The MUSCLES database therefore provides researchers modeling atmospheric photochemistry and escape high-fidelity, $HST$-based, host star SEDs for their calculations.

\section{Targets and Observing Program}

\subsection{MUSCLES Target Stars}  

The MUSCLES target list (Table 1) was chosen to cover a broad range of stellar types (K1V~--~M5V; 4 K dwarfs and 7 M dwarfs), exoplanet masses, and semi-major axes; including most of the known M dwarf exoplanet host stars located within $d$~$\lesssim$~13pc (7/12), while excluding flare stars (e.g. GJ 674) that require intensive multi-wavelength monitoring to clear $HST$ instrument safety protocols.   The M dwarfs span a range of spectral types (from M1~--~M5), a range of X-ray luminosity fractions (log$_{10}$($L_{X}$/$L_{Bol}$)~$\approx$~$-$5.1 to $-$4.4), an indicator of activity level), and planetary systems ranging from Jupiters (GJ 832) to super-Neptunes (GJ~436) to super-Earths (GJ~1214). About $\sim$65~\%  of the exoplanet host stars in our sample (7/11) harbor Super-Earths ($M_{plan}$~$<$~10~$M_{\oplus}$; {\bf bold} in Table~1).  In the brief summaries of the star-planetary systems given in the Appendix, we refer to the {\it M sin i} of the planets as their mass as a shorthand.     

With the exception of $\epsilon$~Eri, the MUSCLES stars are not traditionally classified as active or flare stars (in contrast with widely studied M dwarf flare stars such as AU Mic or AD Leo).  The stars in our sample are considered ``optically inactive'', based on their H$\alpha$ absorption spectra~\citep{gizis02}.  However, the \ion{Ca}{2} H and K emission line cores are a more straightforward means of diagnosing chromospheric activity in the optical~\citep{silva11} because the \ion{Ca}{2} emission line flux increases with activity while H$\alpha$ first shows enhanced absorption before becoming a strong emission line with increasing activity. All of our stars with measured \ion{Ca}{2} H and K profiles show weak but detectable emission (equivalent widths, W$_{\lambda}$(\ion{Ca}{2}) $>$~0; note that the references below present emission lines as positive equivalent widths, opposite from the traditional convention), indicating that at least a low level of chromospheric activity is present in these stars~\citep{rauscher06,walkowicz09}.  In addition, \citet{hawley14} have shown that inactive M dwarfs display flares in their $Kepler$ light curves, confirming that chromospheric activity is present on MUSCLES-type stars.   
Figure 3 shows the \ion{Ca}{2} K line equivalent width as a function of (B~--~V) color for six of the M dwarfs in the sample, overplotted on the data from~\citet{rauscher06}.  The \ion{Ca}{2} K emission is in the range 0.1~$<$~W$_{\lambda}$(\ion{Ca}{2})~$<$~0.8~\AA, approximately an order of magnitude smaller than traditional flares stars EV Lac, AU Mic, AD Leo, and Proxima Cen (also shown in Figure 3).   
According to the M dwarf classification scheme of~\citet{walkowicz09}, these stars have intermediate chromospheres and are referred to as ``weakly active''.  In the Appendix, we present brief descriptions of each of the stars studied here.  For more detailed descriptions of the stellar parameters of the MUSCLES target stars, we refer the reader to~\citet{loyd15} and \citet{youngblood15}.

\begin{figure}
\begin{center}
\epsfig{figure=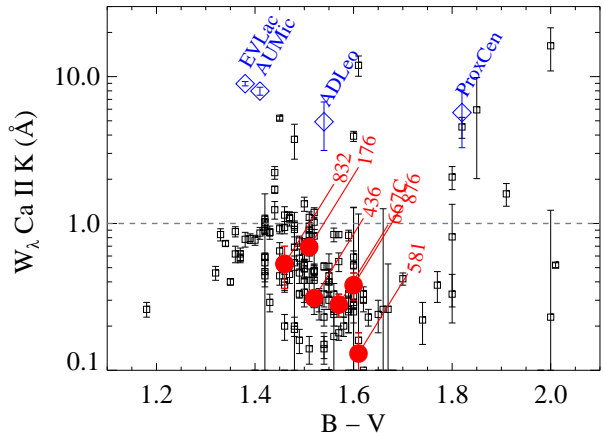,width=2.6in,angle=90}
\vspace{-0.2in}
\caption{
\label{cosovly} The \ion{Ca}{2} K equivalent width as a function of B~--~V color~\citep{walkowicz09,rauscher06} for bright M dwarfs and the MUSCLES M dwarf targets.  The MUSCLES stars (red circles) are characterized as ``weakly active'' with \ion{Ca}{2} activity indicators approximately an order of magnitude lower than traditional M dwarf flare stars (blue diamonds, EV Lac, AU Mic, AD Leo, and Proxima Cen).   Note that large, positive equivalent widths indicate stronger activity (emission line flux) in this presentation. }
\end{center}
\end{figure}

\subsection{MUSCLES Observing Strategy}

{\it Ultraviolet Observations, 100~--~3200~\AA}~--~In order to obtain a full census of the UV emission incident on the habitable zones of low-mass stars, we require $HST$ spectral coverage from 1150~--~3100~\AA: the G130M, G160M, and G230L modes of COS, and the G140M, E140M, E230M, E230H, and G230L modes of STIS provide spectral coverage across this bandpass.   This combination of instrumental settings allows us to catalog the stellar emission lines that are relevant to the photoexcitation of the primary atmospheric constituents of the exoplanets in these systems.   The \ion{H}{1} Ly$\alpha$ emission line dominates the total UV luminosity of M dwarfs~\citep{france12a}.  We use the G140M mode of STIS with the 52\arcsec~$\times$~0.1\arcsec\ slit to measure the Ly$\alpha$ profile for our M dwarf targets.  We have previously demonstrated that this technique can produce high-quality measurements of the local Ly$\alpha$ flux~\citep{france13}.  For the brighter K dwarfs, we employ the STIS E140M mode, using the 0.2\arcsec~$\times$~0.06\arcsec\ slit to resolve the intrinsic line profile and minimize contamination by telluric \ion{H}{1}.   Resonant scattering of Ly$\alpha$ in the local ISM requires that the line must be reconstructed to provide a reliable measure of the intrinsic Ly$\alpha$ radiation field in these exoplanetary systems.    We direct the reader to Paper II for a detailed description of the Ly$\alpha$ reconstruction developed for the MUSCLES Treasury Survey~\citep{youngblood15}.

In the FUV (except Ly$\alpha$) we use COS G130M (\ion{Si}{3} $\lambda$1206, \ion{N}{5} $\lambda$1240, \ion{C}{2} $\lambda$1335, \ion{Si}{4} $\lambda$1400 lines) and COS G160M (\ion{C}{4}$\lambda$1550, \ion{He}{2} $\lambda$1640, \ion{Al}{2} $\lambda$1671 lines).
Emission from Ly$\alpha$, \ion{Si}{4}, \ion{C}{4}, and \ion{He}{2} is particularly interesting because these lines provide constraints on the Lyman continuum/EUV (200~$\lesssim$~$\lambda$~$\lesssim$~900~\AA) irradiance in these systems (Linsky et al. 2014; Shkolnik \& Barman 2014).\nocite{shkolnik14b}   COS is essential for a moderate spectral resolution FUV line census as the lower effective area and higher detector background of STIS make observations of all but the very brightest emission line (\ion{H}{1} Ly$\alpha$) prohibitively time consuming.  We use the medium resolution (``M'') COS modes to resolve chromospheric emission lines and maximize contrast from narrow spectral features.

At NUV wavelengths, we use STIS G230L ($\lambda$~$>$~2200~\AA) to observe the NUV continuum, \ion{Fe}{2} $\lambda$2400 and $\lambda$2600, and \ion{Mg}{2} $\lambda$2800, but take advantage of the superior sensitivity of the COS G230L mode to observe the 1750~--~2200~\AA\ region that is important for the photodissociation of O$_{2}$ and the production of O$_{3}$.   For targets that exceed the G230L bright-object limit (the K stars, $\epsilon$ Eri, HD 40307, HD 85512, and HD 97658), we use the higher-resolution E230M mode ($\lambda$1978 + $\lambda$2707 settings), covering the 1800~--~3100~\AA\ bandpass.  Target brightness limits dictated that we employ the STIS E230H mode for observations of the \ion{Mg}{2} emission lines on the brightest K dwarf ($\epsilon$ Eri).    

\begin{figure*}
\figurenum{4a}
\begin{center}
\epsfig{figure=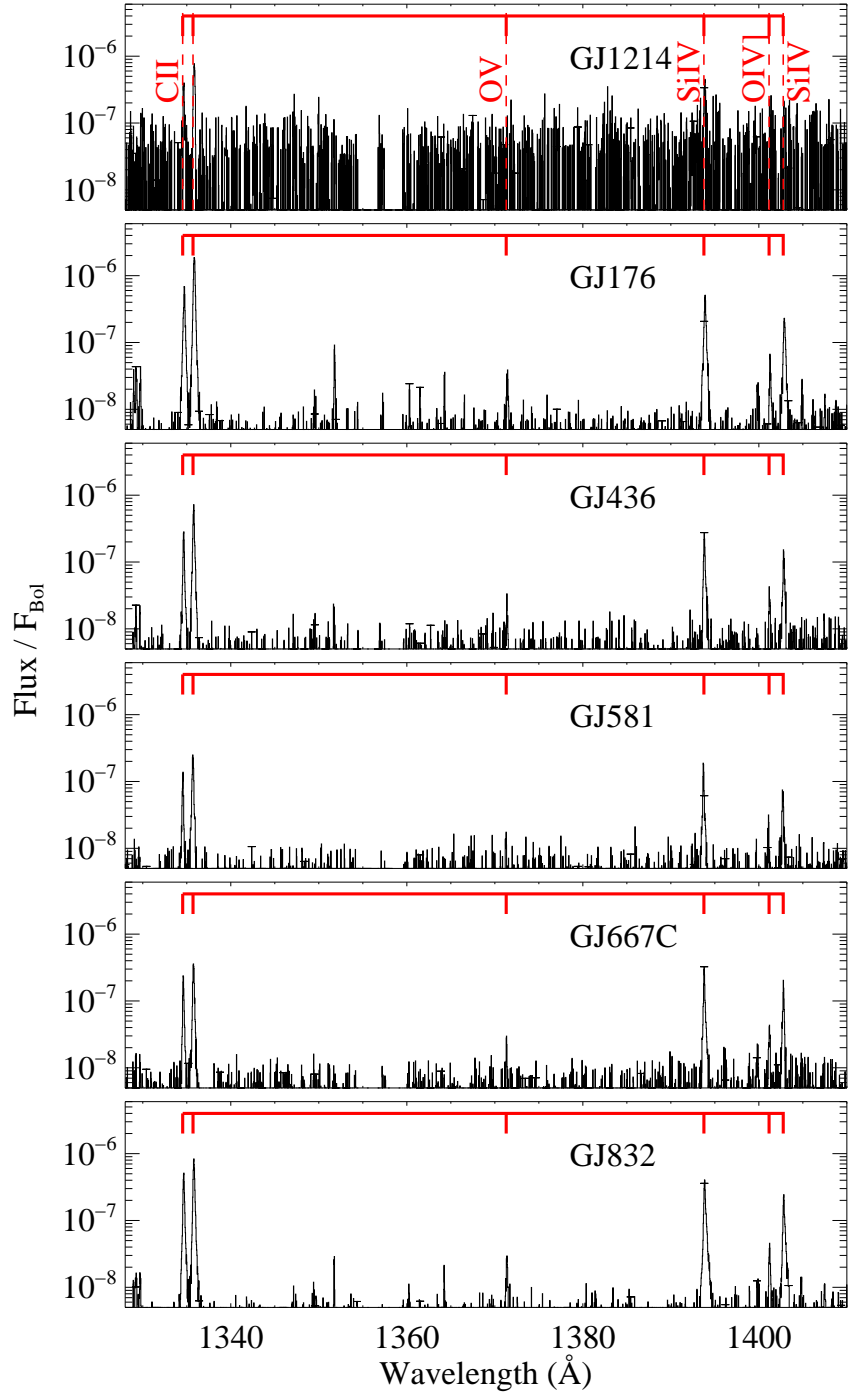,width=4.14in,angle=00}
\vspace{+0.4in}
\caption{
\label{cosovly} A sample of the FUV spectra (1328~--~1410~\AA) of each MUSCLES target.  The flux spectra are normalized by their total bolometric luminosity. These spectra are the coadded observations acquired with the $HST$-COS G130M grating over 5 consecutive orbits, designed to produce both high-S/N FUV spectra and characterize the energetic flare frequency on typical low-mass exoplanet host stars (Section 4.4).  Emission lines formed at temperatures from roughly 10$^{4}$~--~10$^{5}$ K are labeled.  The broad features near 1356 and 1358~\AA\ are geocoronal \ion{O}{1}]. }
\end{center}
\end{figure*}

\begin{figure*}
\figurenum{4b}
\begin{center}
\epsfig{figure=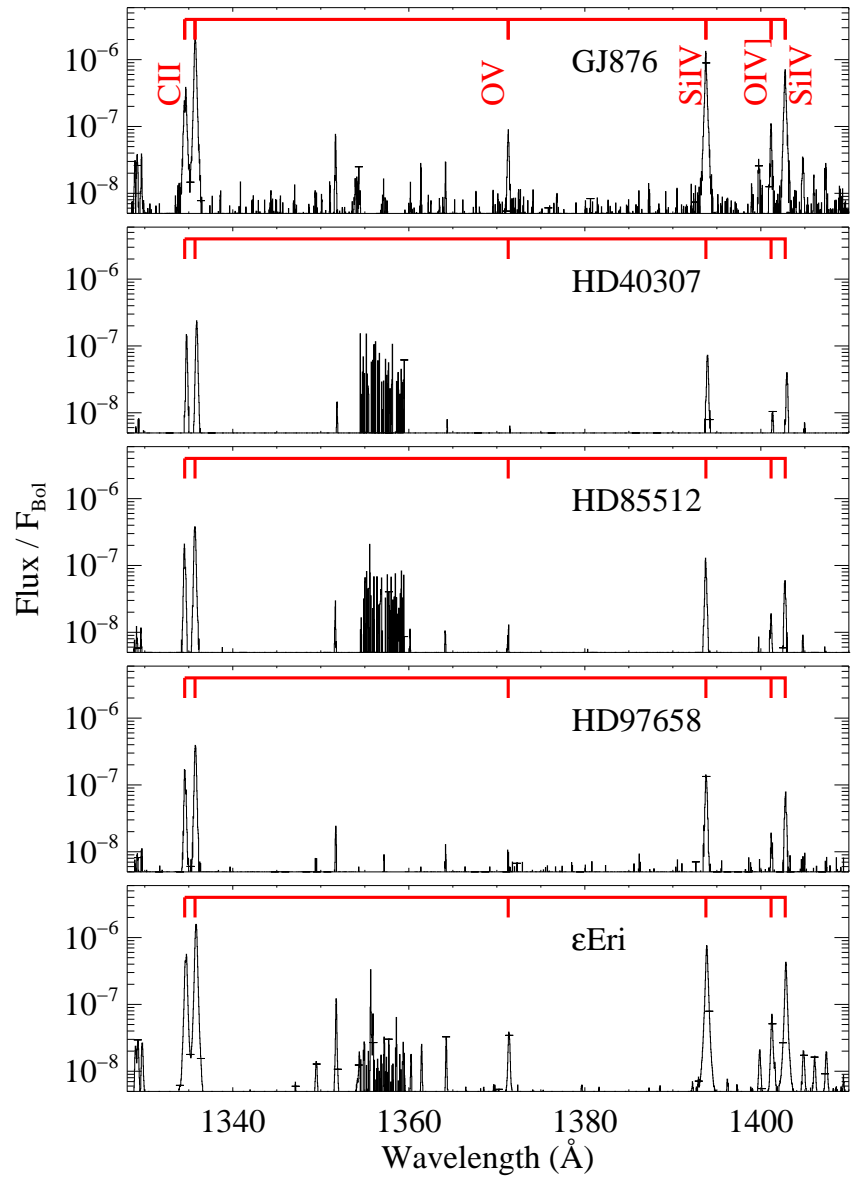,width=4.14in,angle=00}
\vspace{+0.4in}
\caption{
\label{cosovly} same as Figure 4a. }
\end{center}
\end{figure*}

\begin{figure*}
\figurenum{5a}
\begin{center}
\epsfig{figure=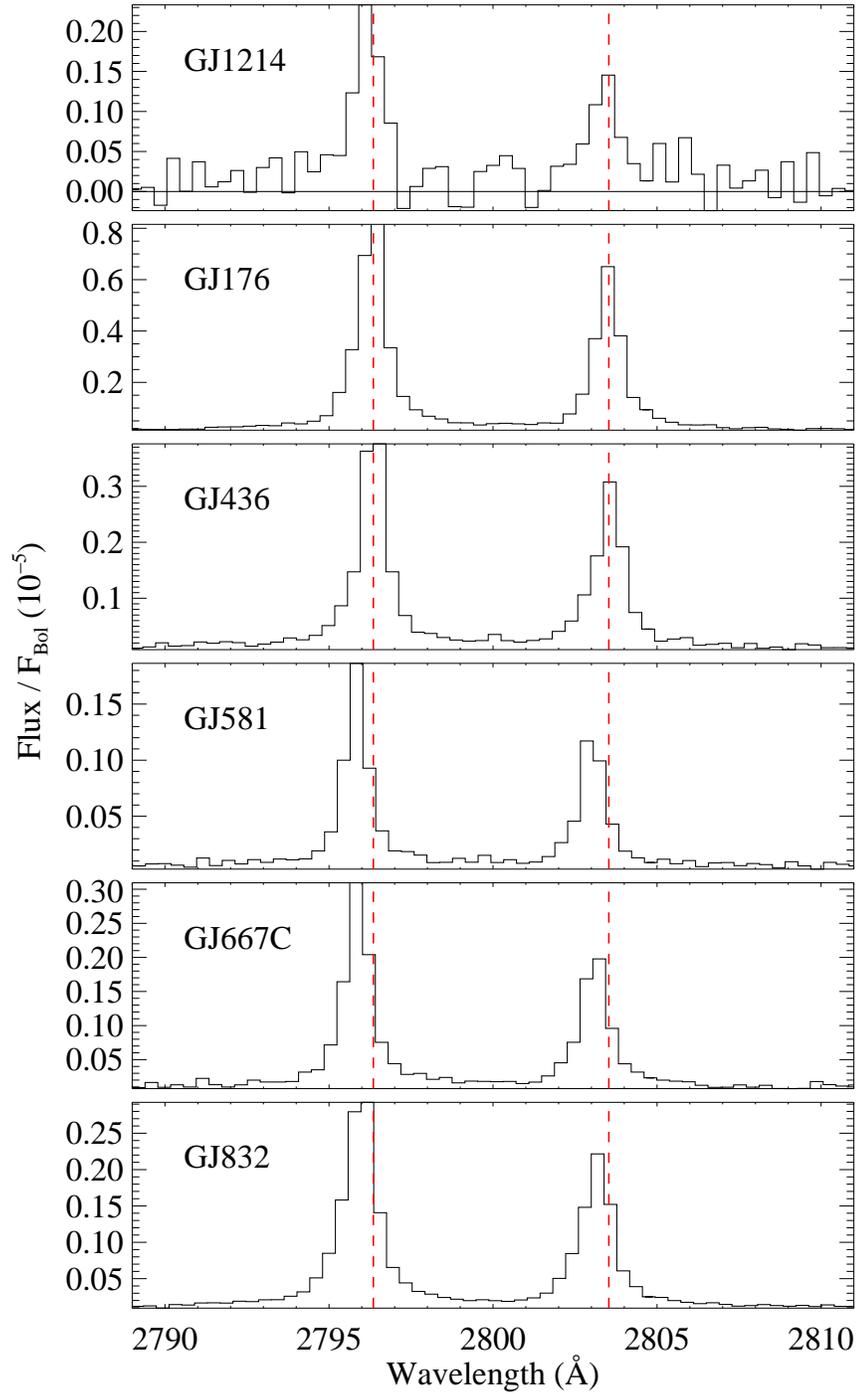,width=4.14in,angle=00}
\vspace{+0.4in}
\caption{
\label{cosovly} NUV spectra (2789~--~2811~\AA) of each of the MUSCLES targets, centered on the \ion{Mg}{2} doublet.  The spectra are normalized by their total bolometric luminosity. These spectra were taken primarily with the COS G230L grating observations in order to show a uniform comparison.  For the brighter K stars, we display higher resolution $HST$-STIS E230M and E230H observations.  At higher resolution, one can resolve the absorption from interstellar Mg$^{+}$ ions. Dashed lines identify the \ion{Mg}{2} rest wavelengths.  }
\end{center}
\end{figure*}

\begin{figure*}
\figurenum{5b}
\begin{center}
\epsfig{figure=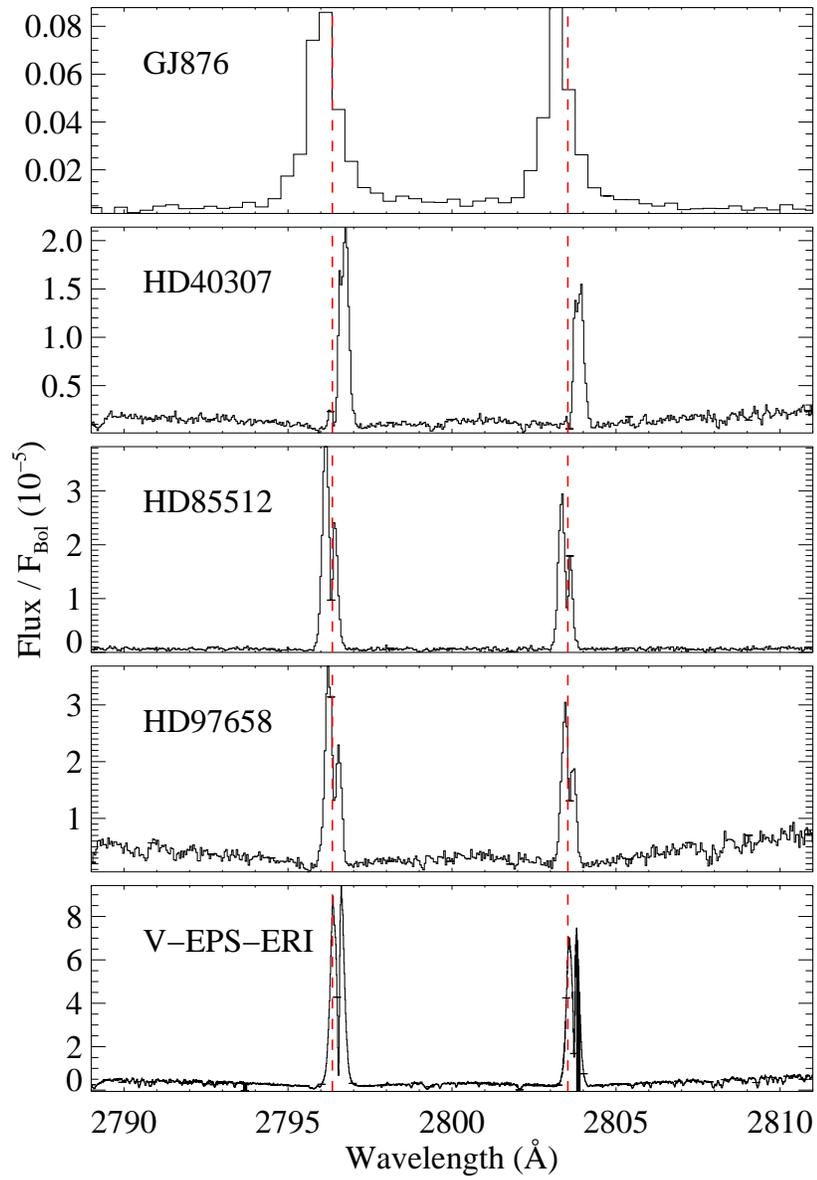,width=4.14in,angle=00}
\vspace{+0.4in}
\caption{
\label{cosovly} same as Figure 5a. }
\end{center}
\end{figure*}

\begin{figure*}
\figurenum{6}
\begin{center}\hspace{-1.0in}
\epsfig{figure=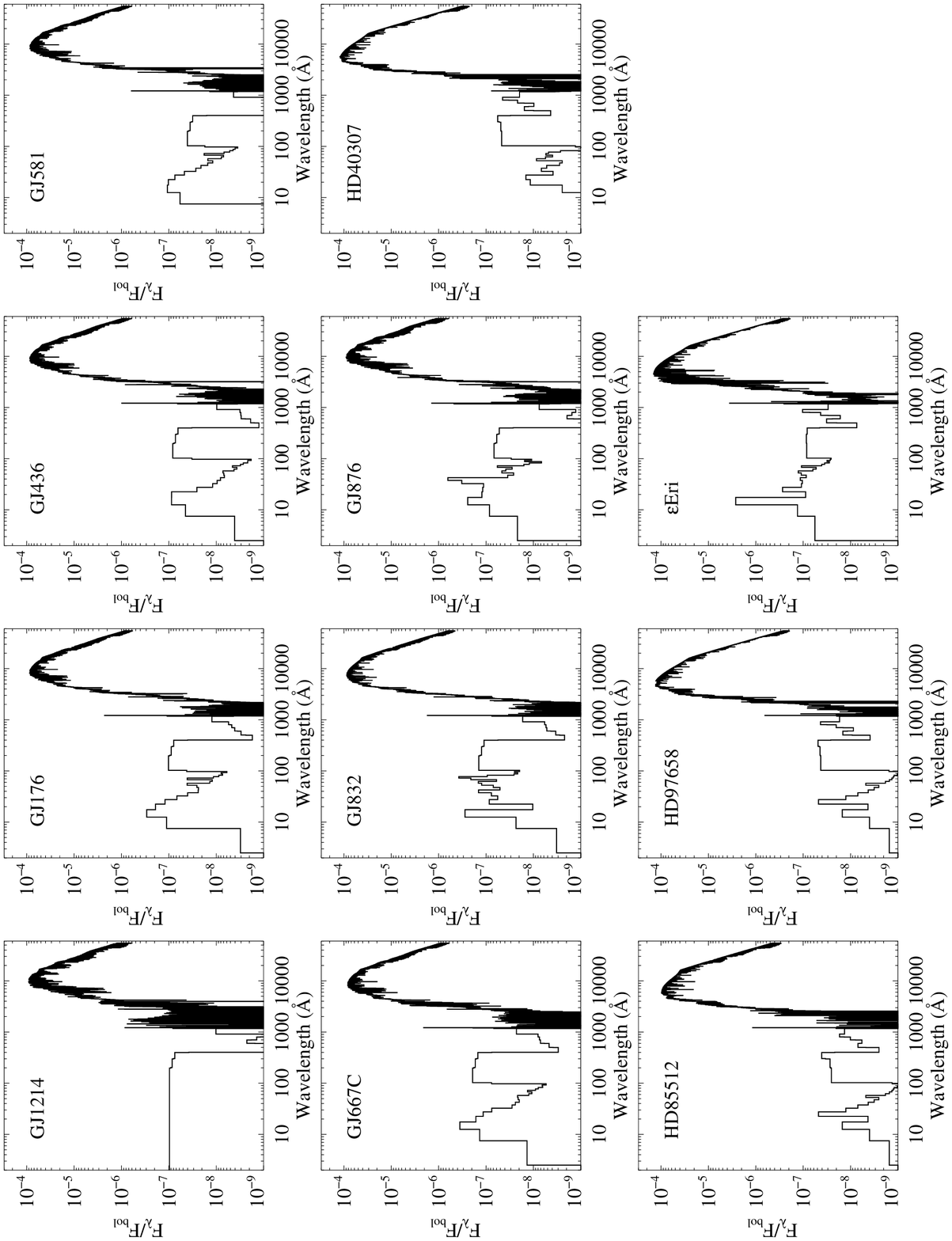,width=5.9in,angle=180}
\vspace{+0.2in}
\caption{
\label{cosovly} Panchromatic SEDs of the MUSCLES targets, binned to 5~\AA\ per pixel for display purposes.  \ion{H}{1} Ly$\alpha$ ($\lambda$~1216~\AA) is observed above the surrounding UV emission and contiuum features.   Both native and 1~\AA\ resolution SEDs are available on the MAST High-Level Science Products site: {\tt https://archive.stsci.edu/prepds/muscles/}.    }
\end{center}
\end{figure*}

The Ly$\alpha$ fluxes are combined with models of solar active regions~\citep{fontenla11} to estimate the EUV luminosity in the wavelength region 100~--~1170~\AA, in 100~\AA\ bins (see Linsky et al. 2014 and Paper II).  
These EUV calculations were compared to the GJ 832 atmosphere model developed by~\citet{fontenla15}; we found that the 10~--~100~\AA\ and 100~--~1000\AA\ bands differed by less than a factor of two.  Given that no inactive M dwarf EUV observations exist as a reference point, we consider this intermodel agreement to sufficient.  
 
 The MUSCLES observing plan was designed to study both the spectral and temporal variability of low-mass exoplanet host stars on the characteristic timescale for UV/optical variability (Kowalski et al. 2009).  Using the most sensitive UV photon-counting mode on $HST$ (COS G130M) and the time-tag capability of the COS microchannel plate detector~\citep{green12}, we measure the 10$^{4-5}$ K chromospheric and transition region activity indicators (using the \ion{C}{2}, \ion{Si}{3}, \ion{Si}{4}, and \ion{N}{5} emission lines) in 8-hour intervals (5 contiguous spacecraft orbits for each star).  Quasi-simultaneous X-ray observations were coordinated with these visits when possible (see below).  While this strategy does not provide continuous coverage due to Earth occultation, it is optimized for constraining the types and frequency of flare behavior on low-mass exoplanet host stars.    Since the characteristic timescale for UV/optical flare activity on M dwarfs is thought to be minutes-to-hours (Welsh et al 2007; Kowalski et al. 2009; Loyd \& France 2014), our observations are ideal for quantifying the importance of flare activity to the local UV radiation field in these systems.  
 
With the above considerations in mind, we arranged the UV~--~visible observations  for each target  into campaigns comprising 3 $HST$ visits: {\bf 1)} COS G130M,  {\bf 2)} COS G160M+G230L,  {\bf 3)} STIS G/E140M+G230L+G430L, all executed within a day of each other to mitigate uncertainties introduced by month or year timescale variations in the stellar flux.   Between 9 and 13 total $HST$ exposures were acquired for each star depending on the target brightness and the observing modes used.  For a graphical description of which modes contribute at which wavelengths for each target, we refer the reader to~\citet{loyd15}.

{\it X-ray Observations, 5~--~50~\AA\ (2.5 keV to 0.25 keV) }~--~We used quasi-simultaneous $Chandra$ and {\it XMM-Newton} observations of the MUSCLES stars to provide temporally consistent SEDs and to explore the wavelength-dependent behavior of stellar flares.   The X-ray observations were coordinated with the COS G130M 5 orbit monitoring program (see above) to establish the longest possible simultaneous baseline over which to explore the panchromatic properties of M and K dwarf flares.  We observed GJ 667C, GJ 436, GJ 176, and GJ 876 using the $Chandra$ ACIS-S back-illuminated S3 chip.   Owing to an $HST$ safing event in 2015 June, the GJ 876 campaign was not simultaneous.   We observed $\epsilon$ Eri and GJ832 with {\it XMM-Newton} because
they are optically bright. $\epsilon$ Eri is by far the brightest X-ray source in our sample and provided a high-quality RGS grating spectrum. $\epsilon$ Eri required use of the pn/MOS “Thick” filters, while GJ832 used the “Medium” filters. We also obtained photometry with the UVM2 filter for GJ 832.  HD 85512 and HD 40307 were observed as part of a complementary {\it XMM-Newton} program (PI~--~A. Brown) using the same configuration as for GJ 832.   For the stars without contemporaneous X-ray data, we obtained spectra from the $Chandra$ and {\it XMM-Newton} archives (Loyd et al. 2016; Brown et al. 2016~--~in prep.).   The coronal model fit to the X-ray data is extended into the 50~--~100~\AA\ region where no direct observations are available.  

{\it Optical Observations, 3200~--~6000~\AA}~--~We carried out complementary ground-based observations as close in time as possible to the MUSCLES UV/X-ray observations.   Our primary optical spectra came from the ARCES and DIS instruments on the Astrophysical Research Consortium (ARC) 3.5-m telescope at Apache Point Observatory (3700~--~10000~\AA, depending on the mode), the 2.15-meter Jorge Sahade telescope at the Complejo Astronomico El Leoncito~\citep{Cincunegui04}, and the FLOYDS intrument on the Las Cumbras Observatory Global Telescope Network~\citep{brown13}.  Multi-band optical photometry was acquired with the Apache Point Observatory 0.5-m ARCSAT telescope (with the Flarecam instrument) and the LCOGT.   The optical data and their correlation with spectral and temporal behavior of the high-energy emission will be presented in a future work by the MUSCLES team~\citep{youngblood16}.  In order to calibrate the UV data with respect to visible/IR photospheric models and the ground-based spectra, we also acquired short optical observations with STIS G430L or G430M (depending on target brightness) during each $HST$ campaign.    

{\it Infrared Extension, $>$~6000~\AA}~--~Stellar atmosphere models considering only the photosphere (e.g., the PHOENIX models) are able to reproduce the peak and Rayleigh-Jeans tail of the stellar SED given the correct prescription for the effective temperature and abundances of atoms and molecules~\citep{allard95,husser13}.  We employ PHOENIX models, matched to the Tycho B and V band fluxes, to calibrate the optical spectroscopy from $HST$-STIS and fill out the panchromatic SEDs from the red-optical to the mid-IR ($\sim$~0.6~--~5.5~$\mu$m; Loyd et al. 2016).  The red-optical and infrared fluxes are important as they make up the majority of the bolometric luminosity from cool stars, regulate the effective surface temperature of orbiting planets, and provide the reference for the activity indicators (e.g., $L$(\ion{N}{5})/$L$$_{Bol}$) presented in this work.


\section{Results:  5~\AA\ to 5~$\mu$m Irradiance Spectra}

The primary goal of the MUSCLES Treasury Survey is to develop a uniform set of irradiance spectra to support the study of extrasolar planets orbiting low-mass stars.  The X-ray through optical observations are described above, a description of the analysis required to complete ``missing'' parts of the SED (intrinsic Ly$\alpha$ line profiles and EUV fluxes) is presented in Paper II~\citep{youngblood15}, and a detailed discussion of the creation of the panchromatic radiation fields is given in Paper III of this series~\citep{loyd15}.   In the following subsections, we present both quantitative and qualitative descriptions of the SEDs, including the behavior of the broadband fluxes as a function of stellar effective temperature and habitable zone location, the distribution of stellar emission lines observed in the $HST$ spectra, the use of individual spectral tracers to estimate the broadband energetic radiation fluxes, and an initial description of the temporal variability of the MUSCLES Treasury dataset.  

For the results presented below, we used both broadband and emission line fluxes.  The broadband fluxes are simply integrals of the panchromatic SED over the wavelength region of interest.  Ly$\alpha$ fluxes are the wavelength-integrated reconstructed line-profiles obtained by~\citet{youngblood15}.  We measured the metal emission lines by employing a Gaussian line-fitting code that takes into account the line-spread function of the instrument used to create that portion of the spectrum.  For instance, \ion{N}{5} and \ion{C}{4} lines are fit with a single-component Gaussian emission line convolved with the appropriate wavelength-dependent $HST$-COS line-spread function\footnote{The COS line-spread function experiences a wavelength dependent non-Gaussianity due to mid-frequency wave-front errors produced by the polishing errors on the $HST$ primary and secondary mirrors; {\tt http://www.stsci.edu/hst/cos/documents/isrs/}}~\citep{france12b}.  Some emission lines are clearly better described by a two-component model~\citep{wood97}, but this only applies to a small number of emission lines in the sample and we assume a single Gaussian for the results presented in this paper.   \ion{Mg}{2} emission line fluxes assume an unaltered Gaussian line-shape for both COS and STIS observations.  As most of the \ion{Mg}{2} observations were made with the low-resolution G230L modes of COS and STIS, an interstellar absorption correction is not possible for all targets, so the \ion{Mg}{2} line fluxes are presented as-measured.      Figures 4 and 5 show sample FUV (1328~--~1410~\AA) and NUV (2789~--~2811~\AA) spectral regions at full resolution with the prominent emission lines labeled.


\subsection{Broadband Fluxes}

Figure 6 shows a montage of the full 5~\AA~--~5~$\mu$m SEDs for the 11 MUSCLES exoplanet host stars, binned to 5~\AA\ pixel$^{-1}$ for display and divided by the bolometric luminosity.  The bolometric luminosity used in this work is simply the integral of the complete SED for each star (including a calculated Rayleigh-Jeans tail at $\lambda$~$>$~5.5~$\mu$m; Loyd et al. 2016).  Many of the component spectra have spectral resolutions much higher than this (resolving powers of $R$~$>$~15,000 for all of the FUV data and up to $R$~=~114,000 for STIS E230H observations of $\epsilon$~Eri), and both full resolution and binned data sets are available on the MAST HLSP archive\footnote{ \tt https://archive.stsci.edu/prepds/muscles/}.  As discussed in the introduction, the broadband behavior of the stellar SED can give general insights into the relevant heating and photochemical rates for all types of exoplanet atmospheres.  

We define the FUV flux, $F$(FUV), as the total stellar flux integrated over the 912~--~1700~\AA\ bandpass, including the reconstructed Ly$\alpha$ emission line, FUV = ($F$(Ly$\alpha$) + $F$(1170~--~1210 + 1220~--~1700) + $F$(912~--~1170)).   $F$(Ly$\alpha$) is the reconstructed Ly$\alpha$ line flux, $F$(1170~--~1210 + 1220~--~1700) is the non-Ly$\alpha$ flux directly measured by $HST$ COS and STIS, and $F$(912~--~1170) is the ``Lyman Ultraviolet'' emission constructed from the solar active region relations for low-mass stars presented by~\citet{linsky14}.  Ly$\alpha$ contributes on average 83\% ($\pm$~5\%) to the total FUV flux.   $F$(NUV) are the combined $HST$ spectra integrated over 1700~--~3200~\AA.  The fractional luminosities in each band as well as key emission lines are given in Table 3.

Figure 7 shows the FUV/NUV flux ratio as a function of stellar effective temperature.  The FUV/NUV flux ratio is in the range 0.5~--~0.7 for the latest M stars (GJ 1214 and GJ 876), is in the range 0.2~--~0.4 for M1 to M3 stars, declines to 0.04 by mid-K, and is $\lesssim$~0.01 for spectral types K2 and earlier.  \citet{france13} have previously shown that the FUV/NUV ratio drops to $\sim$~10$^{-3}$ for solar-type stars.  The trend in FUV/NUV with effective temperature is largely driven by the large increase in photospheric emission moving into the NUV spectral bandpass with increasing effective temperature.  Large FUV fluxes dissociate O$_{2}$ to generate O$_{3}$.  The photospheric NUV photons destroy photochemically produced ozone and keep the atmospheric O$_{3}$ mixing ratios low on Earth-like planets around solar-type stars, in the absence of a disequilibrium process such as life.  

The XUV flux (5~--~911~\AA), the combination of the observed and modeled soft X-ray flux (usually 2.5 keV~--~0.125 keV, or $\approx$~5~--~100~\AA\footnote{We do not include hard X-rays in our panchromatic SEDs as the observations are not easily obtainable and the photoionization cross-sections for most of the relevant atmospheric constituents decline at energies higher than EUV~+~soft-X-rays}) and the calculated EUV flux (100~--~911~\AA), is an important heating agent on all types of planets. XUV irradiance is particularly important for short-period planets~\citep{lammer09}.  
Figure 8 shows the relative XUV and FUV luminosities as a function of effective temperature for the MUSCLES stars.  The total XUV and FUV luminosities are shown to be well-correlated with stellar effective temperature (and therefore stellar mass) and the fractional XUV and FUV luminosities are in the range ~10$^{-5}$~$\lesssim$~$L$(band)/$L_{bol}$~$\lesssim$~10$^{-4}$ with no dependence on stellar effective temperature.  For comparison, the disk-integrated quiet Sun~\citep{woods09} has $L$(XUV)/$L_{Bol}$~=~2~$\times$~10$^{-6}$ and $L$(FUV)/$L_{Bol}$~=~1~$\times$~10$^{-5}$, respectively.  Note however the solar $L$(FUV) contains a contribution from the photosphere, whereas the photospheric contribution in negligible for M and K dwarfs.  From the perspective of integrated planetary atmosphere mass-loss over time, our measurements represent only a conservative lower-limit as the stellar XUV radiation was likely factors of 10~--~100 times higher during the star's younger, more active periods.  The active periods of M stars are more prolonged than the equivalent ``youthful magnetic exuberance''~\citep{ayres10b} period of solar-type stars (e.g., West et al. 2008).  Additionally, it is likely that the stellar mass-loss rates are higher earlier in their evolution~\citep{wood05b}, increasing the potential planetary mass-loss.\nocite{west08}  

We can combine these results into a comprehensive picture of the energetic radiation environment in the HZs around the low-mass stars.  Figure 9 shows the FUV/NUV ratios, the total FUV fluxes, and the total XUV fluxes at the habitable zone orbital distances for each of our targets.  The plots show the average habitable zone distance for each star, $\langle$$r_{HZ}$$\rangle$, computed as the mean of ``runaway greenhouse'' and ``maximum greenhouse'' limits to the HZ presented by~\citet{kopparapu14}.  The error bars on $\langle$$r_{HZ}$$\rangle$ represent these extrema.   One observes the two orders-of-magnitude decline of the FUV/NUV ratio from 0.1~--~0.7~AU, mainly driven by the stellar effective temperature dependence described above.  The FUV HZ fluxes show a weak trend of increasing flux with increasing $r_{HZ}$, however both the M dwarf and K dwarf samples have a factor of roughly 5 dispersion at a given HZ distance.  
For example, while it is possible for the absolute FUV flux to be a factor of 10 greater at 0.7 AU than 0.15 AU, it is also possible that the FUV flux at 0.15 AU is greater than at 0.7 AU.  
A similar dispersion is seen in the XUV fluxes, and there is no statistically significant change in the XUV flux across the habitable zone (Table 2).  Therefore, the average FUV and XUV fluxes in the HZs of M and K dwarfs are 10~--~70 erg cm$^{-2}$ s$^{-1}$.   
We emphasize that the activity level of the individual star must be considered, and direct observations are preferable when available.  In Section 4.3, we will discuss the correlation of the broadband fluxes with specific emission line measurements to simplify the characterization of the broadband fluxes from single emission line flux measurements.   

\begin{figure}
\figurenum{7}
\begin{center}
\epsfig{figure=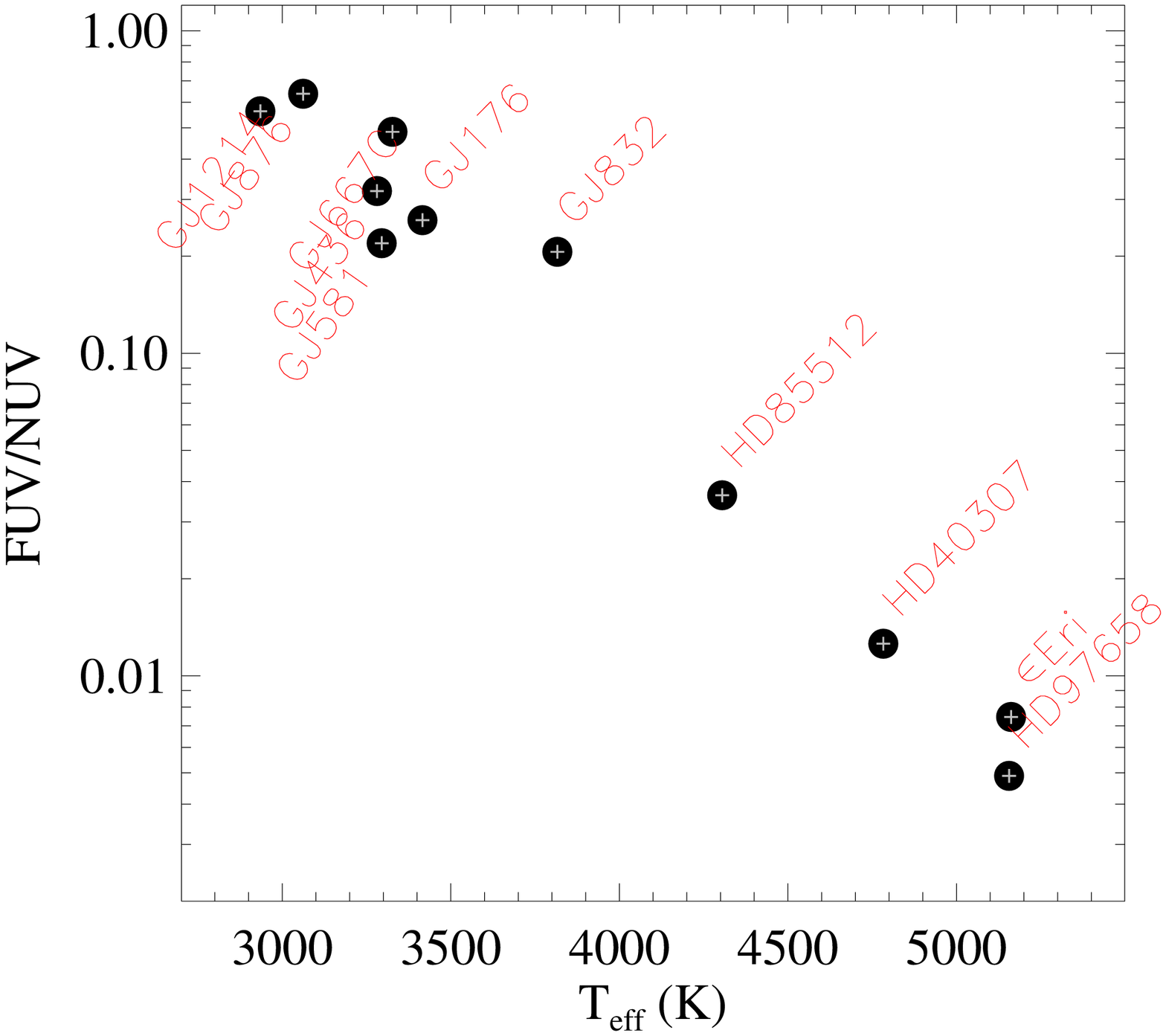,width=2.6in,angle=90}
\vspace{+0.08in}
\caption{
\label{cosovly} The FUV/NUV (912~--~1700~\AA)/(1700~--~3200~\AA) stellar flux ratio decreases with T$_{eff}$ by a factor of $\sim$~100 from mid-M dwarfs to early-K dwarfs. This relationship is largely driven by the increasing photospheric flux in the NUV band of hotter stars.  The FUV/NUV flux ratio has been shown to impact oxygen chemistry on Earth-like planets in the HZ~\citep{hu12,tian14}.  }
\end{center}
\end{figure}

\begin{figure}
\figurenum{8}
\begin{center}
\epsfig{figure=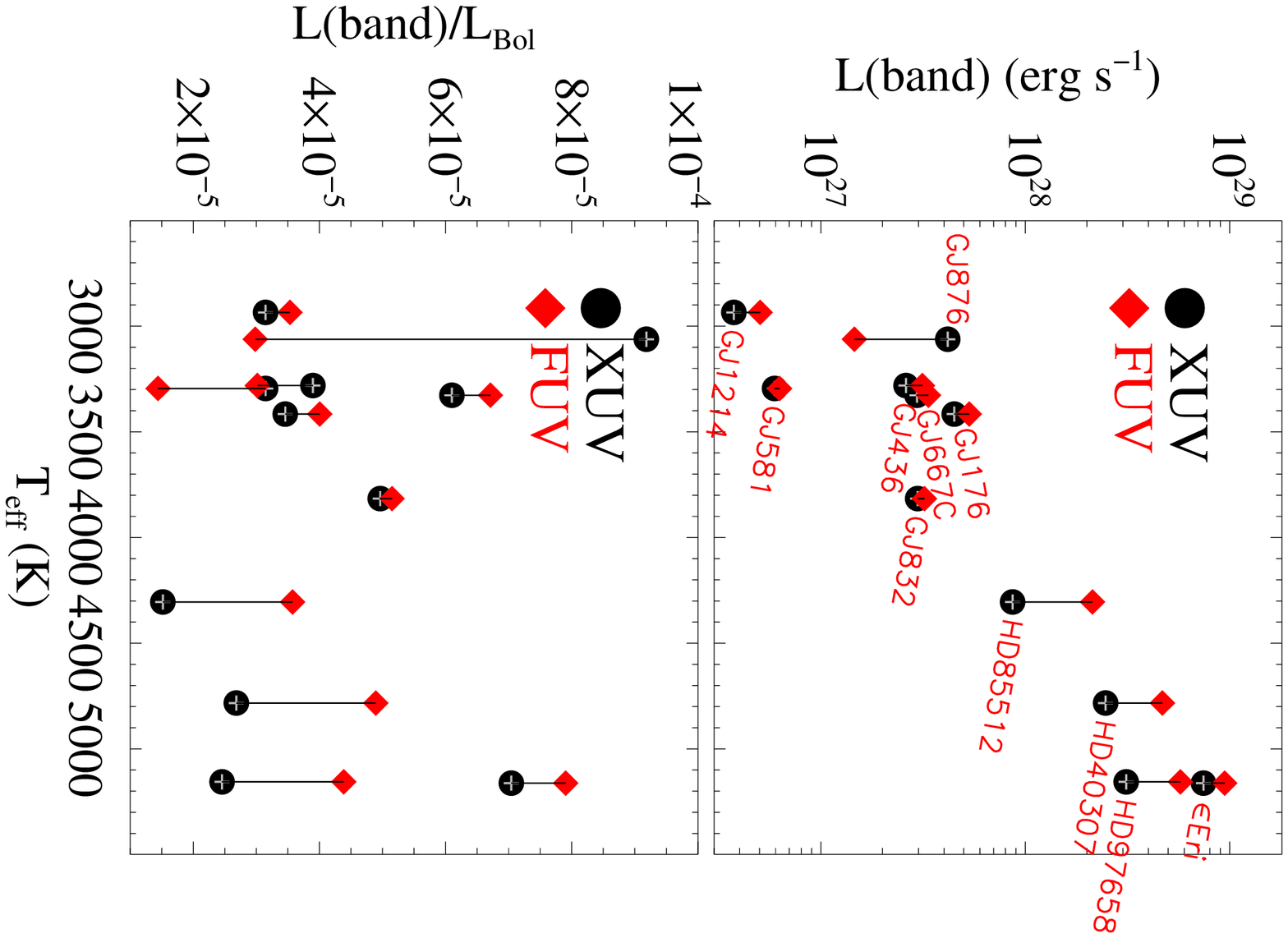,width=3.2in,angle=90}
\vspace{+0.2in}
\caption{
\label{cosovly} The total UV luminosity is proportional to the effective temperature (and therefore stellar mass), whereas the fractional luminosity is roughly constant across the sample.  ($top$)  The total XUV (5~\AA~--~911~\AA) luminosity and total FUV (912~--~1700~\AA) luminosity increase with stellar effective temperature.  ($bottom$) The fractional luminosities in the XUV and FUV bands are in the range 10$^{-5}$~--~10$^{-4}$ for all of the MUSCLES targets.  The black circles represent the XUV luminosity while the red diamonds represent the FUV luminosity. }
\end{center}
\end{figure}

\subsection{1150~--~3200~\AA\ Stellar Emission Lines: The Ubiquity of UV Emission }

Chromospheric and transition region emission lines are observed in $all$ of the MUSCLES spectra, arguing that all exoplanet host stars with spectral type M6 and earlier have UV-active atmospheres (Figures 4 and 5).  This seems to rule out photosphere-only models of cool stars and indicates that the chromospheric, transition region, and coronal emission must be included for accurate modeling of the atmospheres of planets orbiting these stars.   Bright emission lines in the MUSCLES spectra with chromospheric formation temperatures in the range $\sim$~(4~--~30)~$\times$~10$^{3}$ K include \ion{H}{1} Ly$\alpha$, \ion{Si}{2} $\lambda$1264, 1526, 1808, \ion{C}{2} $\lambda$$\lambda$1334,1335, \ion{Al}{2} $\lambda$1671, \ion{Fe}{2} multiplets near 2400 and 2600~\AA, and \ion{Mg}{2}~$\lambda$$\lambda$2796,2803 (see e.g., the M dwarf contribution functions presented by Fontenla et al. 2015).  We observe many transition region lines with formation temperatures from $\sim$~(40~--~200)~$\times$~10$^{3}$ K, including the \ion{C}{3} 1175 multiplet, \ion{Si}{3} $\lambda$1206, \ion{O}{5} $\lambda$1218, 1371, \ion{N}{5} $\lambda$$\lambda$1239,1243, \ion{O}{4}] $\lambda$1401, \ion{Si}{4} $\lambda$$\lambda$1394,1403, \ion{C}{4} $\lambda$$\lambda$1548,1550, and \ion{He}{2} $\lambda$1640.  The coronal iron lines, \ion{Fe}{12} $\lambda$1242 and \ion{Fe}{21} $\lambda$1354~\citep{ayres03}, are observed in a fraction of $HST$-COS observations (7/11 stars show \ion{Fe}{12} emission lines while only GJ 832 and GJ 876 show \ion{Fe}{21} emission lines).  Taken together with the X-ray observations, these highly ionized iron lines demonstrate the presence of coronal gas in the atmospheres of all of our stars and enable an alternative calculation of the EUV irradiance using emission measure techniques~\citep{forcada11,chadney15}.   We will present a detailed discussion of the atmospheric kinematics derived from emission line parameters in an upcoming work (Linsky et al.~--~in preparation).    

A major complication in previous attempts to assemble panchromatic radiation fields, particularly of M dwarfs, is the flux variability between observations separated by years (e.g., comparing $IUE$ spectra from the 1980s with $ROSAT$ X-ray data from the 1990s and $HST$ observations of Ly$\alpha$ from the 2000s, Linsky et al. 2013, 2014).\nocite{linsky13,linsky14}  One of the goals of the MUSCLES Treasury Survey is to obtain multi-wavelength observations close in time to minimize this large systematic uncertainty.  In the MUSCLES observing strategy, emission lines with similar formation temperatures were acquired on different $HST$ visits (owing to their inclusion on different $HST$ grating modes) separated by 18~--~48 hours.  The overlap in formation temperature and spectral coverage between adjacent modes facilitates scaling over calibration variations~\citep{loyd15} and smaller day-to-day variations.  

This approach has been successful~--~Figure 10 shows the relationship of the fractional luminosity in the transition region ions \ion{N}{5} and \ion{C}{4}.  These lines have formation temperatures within a factor of two of each other (between (1~--~2)~$\times$~10$^{5}$~K) and have been shown to be tightly correlated in numerous astrophysical plasmas, including the atmospheres and accretion columns around young stars~\citep{oranje86,ardila13,france14}.     The \ion{N}{5} versus \ion{C}{4} correlation is very well maintained over the MUSCLES sample despite the non-simultaneous observations, with a Pearson correlation coefficient of 0.90 ($\rho$ in the legend of Figure 10) and a probability of no correlation of 6.0~$\times$~10$^{-5}$ ($n$ in the legend of Figure 10).  Because we do not exclude discrete impulsive flares, the MUSCLES spectra can be considered to be an accurate snapshot of the average stellar spectrum (averaged on the timescale of hours-to-days).   Variability on the timescales of the solar cycle (or stellar cycles, years) will be addressed by developing XUV~--~optical tracer correlations and carrying out long-term monitoring from ground-based facilities~(Section 1.3).   

The emission lines in the MUSCLES stars also show an evolution of decreasing fractional luminosity of the transition region with increasing effective temperature.  Figure 11 ($top$) shows the fractional \ion{N}{5} luminosity as a function of effective temperature.  
Given that the total XUV and FUV factional luminosities are approximately constant with effective temperature (Figure 8, $lower$), this suggests that the upper transition region activity (traced by~\ion{N}{5}) declines faster than the rest of the upper atmosphere (e.g., the chromosphere and corona), possibly relating to the pressure-density structure of the stellar atmosphere changing with mass~\citep{fontenla15}.  While the Pearson correlation coefficient is only $-$0.60 ($n$~=~3.5~$\times$~10$^{-2}$), this is skewed by the inclusion of $\epsilon$~Eri, the only active star in the sample.  Excluding $\epsilon$~Eri, the coefficient becomes $-$0.65 ($n$~=~9.1~$\times$~10$^{-3}$).    The middle plot in Figure 11 shows the total \ion{N}{5} luminosity with the stellar rotation rate, suggesting a period-activity relation analogous to the well-studied relationship in solar-type stars~(Ribas et al. 2005; see also Engle \& Guinan 2011 for M dwarf X-ray evolution).~\nocite{ribas05,engle11}   The notable exception is GJ 876, with a \ion{N}{5} luminosity approximately an order of magnitude larger than expected based on its rotational period.  This is partially the result of the strong flare activity on this star (see Section 4.4), and also suggests that the period is not well-determined for this star.  Excluding GJ 876, the Pearson and Spearman coefficients are $-$0.62 and $-$0.92, respectively.   If GJ 876 fell on this trend, we would expect a rotation period closer to $\sim$~40 days.  Interestingly, there is no trend in the fractional \ion{N}{5} luminosity with stellar rotation rate (Figure 11, $bottom$).    

\begin{figure*}
\figurenum{9}
\begin{center}\hspace{-1.0in}
\epsfig{figure=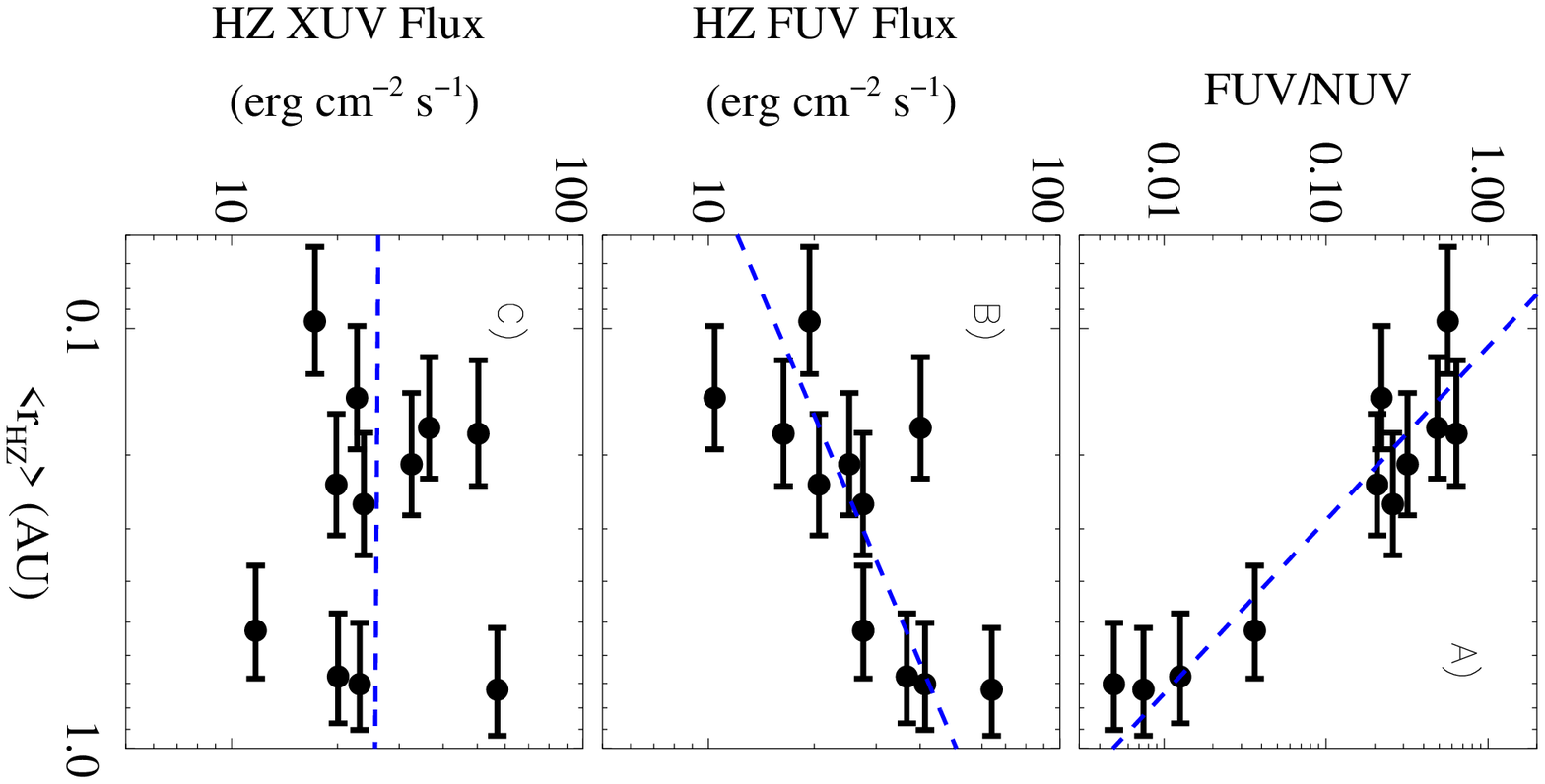,width=6.6in,angle=90}
\vspace{+0.4in}
\caption{
\label{cosovly} The evolution of the ultraviolet radiation environment over the habitable zone for low-mass stars.  The habitable zone distance, $\langle$$r_{HZ}$$\rangle$, is defined as the average of the `'runaway greenhouse'' and ``maximum greenhouse'' limits~\citep{kopparapu14}, and the error bars show these extrema.  
The FUV/NUV ratio increases by a factor of 100 from 0.7 AU to 0.1 AU (panel A), while the average absolute FUV (panel B) flux decrease by factors of three to five towards the inner habitable zone. Note that the scatter in the FUV flux at any given HZ radius is at least as large as the correlations indicated by the log-log fits shown as the blue dashed lines.   The average absolute EUV (panel C) flux is constant across the HZ.     }
\end{center}
\end{figure*}

\begin{figure}
\figurenum{10}
\begin{center}
\epsfig{figure=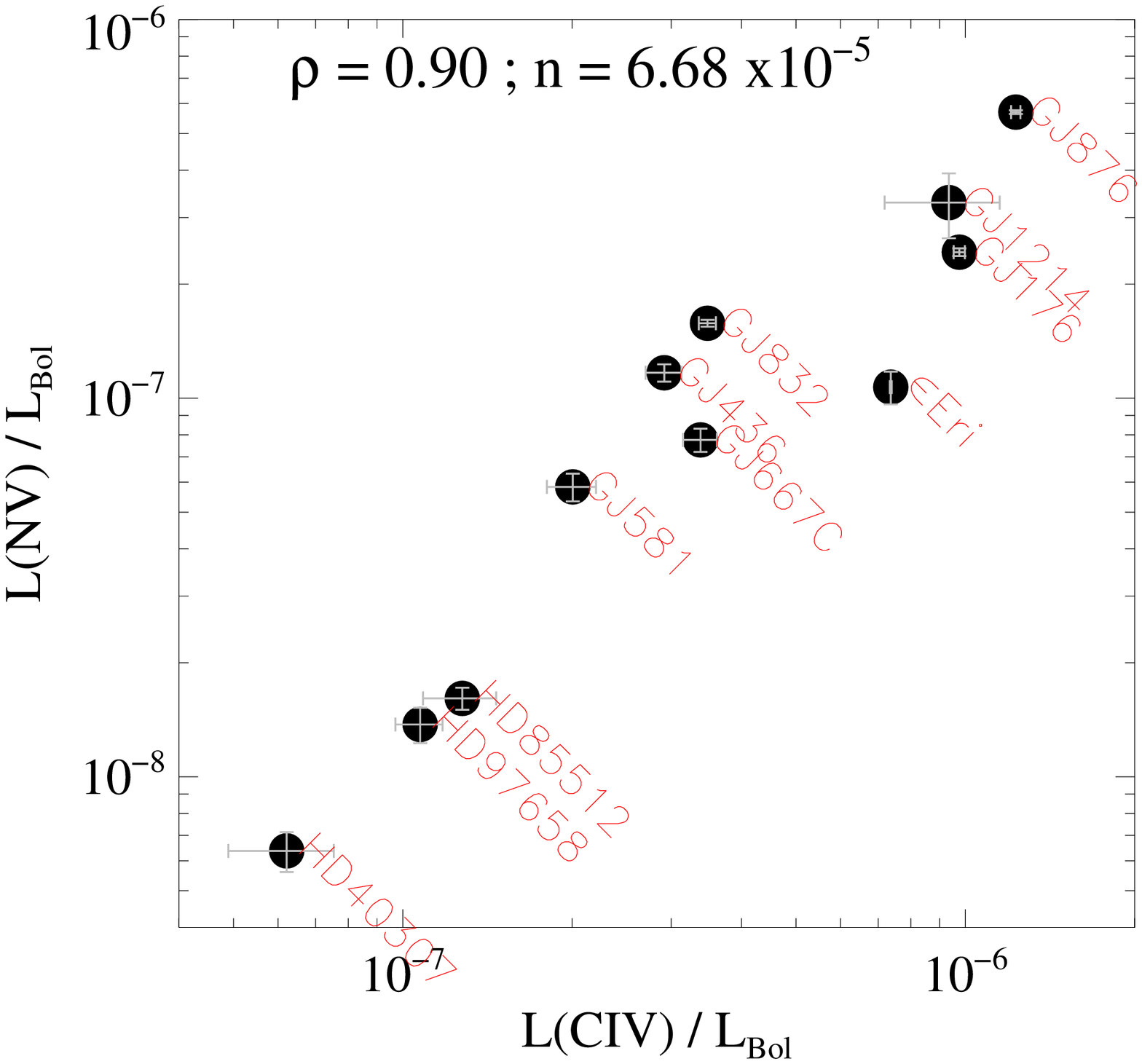,width=2.6in,angle=90}
\vspace{+0.1in}
\caption{
\label{cosovly} A comparison of the fractional \ion{C}{4} and \ion{N}{5} luminosities (relative to the bolometric luminosity) of the MUSCLES samples.  The Pearson correlation coefficient ($\rho$) and a statistical measure of the possibility of no correlation ($n$) are shown at the top of the panel, quantifying the obvious strong correlation between the two emission lines.  
This plot shows that the observation-averaged stellar fluxes from similar formation temperature ranges do not vary significantly on the $\sim$~1 day time-scales separating the MUSCLES observations.  
}
\end{center}
\end{figure}

\begin{figure}
\figurenum{11}
\begin{center}
\epsfig{figure=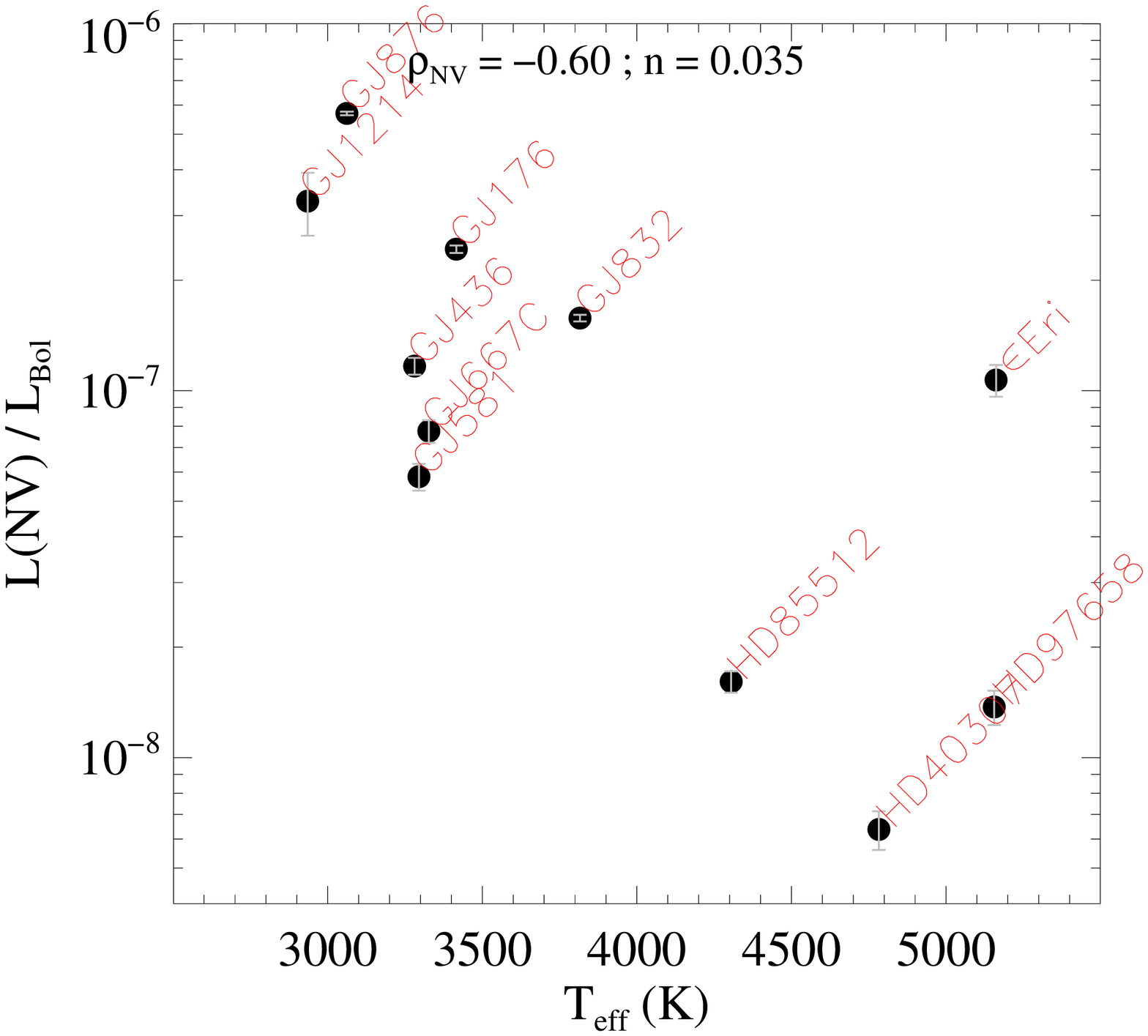,width=2.52in,angle=90}
\epsfig{figure=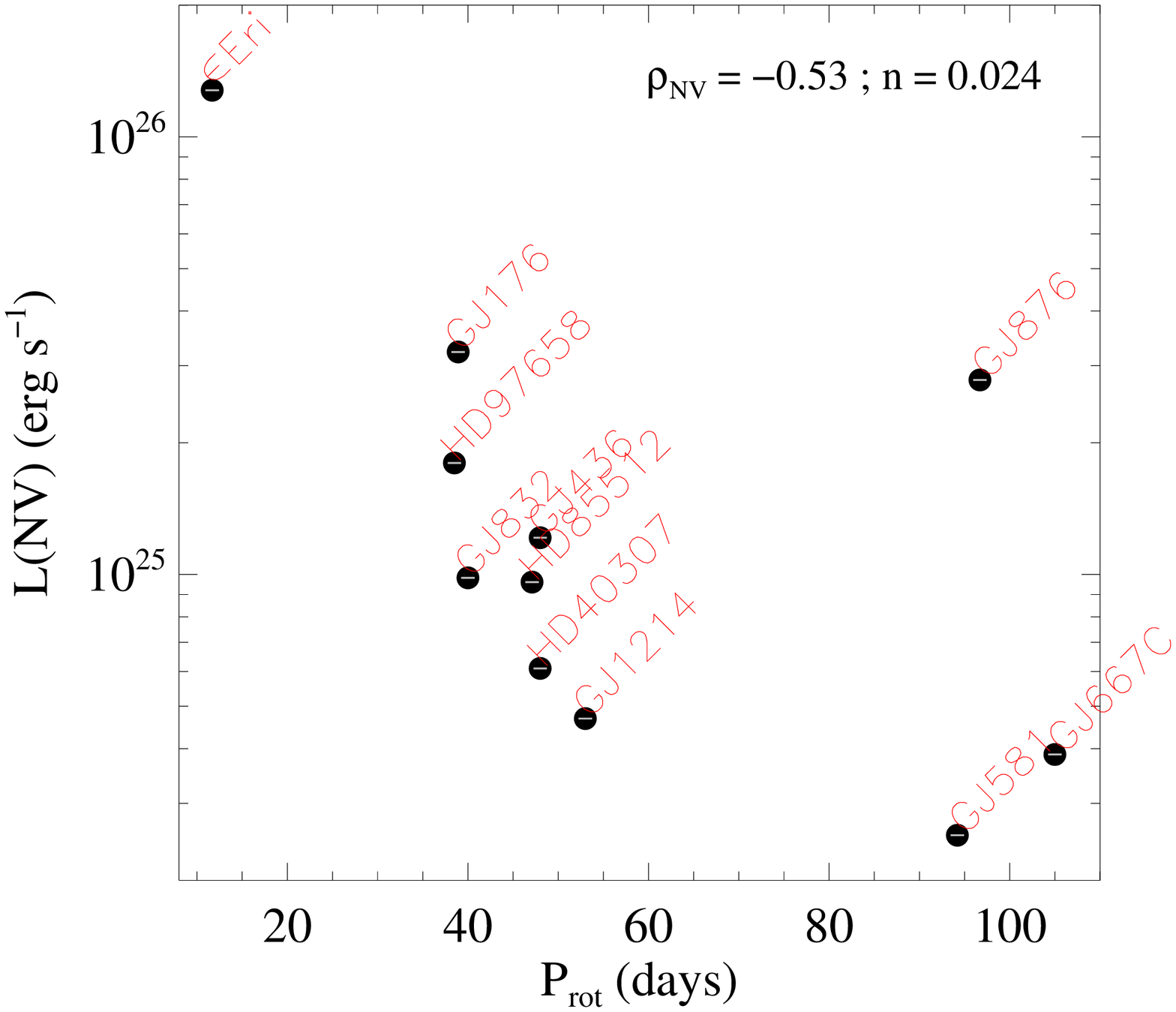,width=2.52in,angle=90}
\epsfig{figure=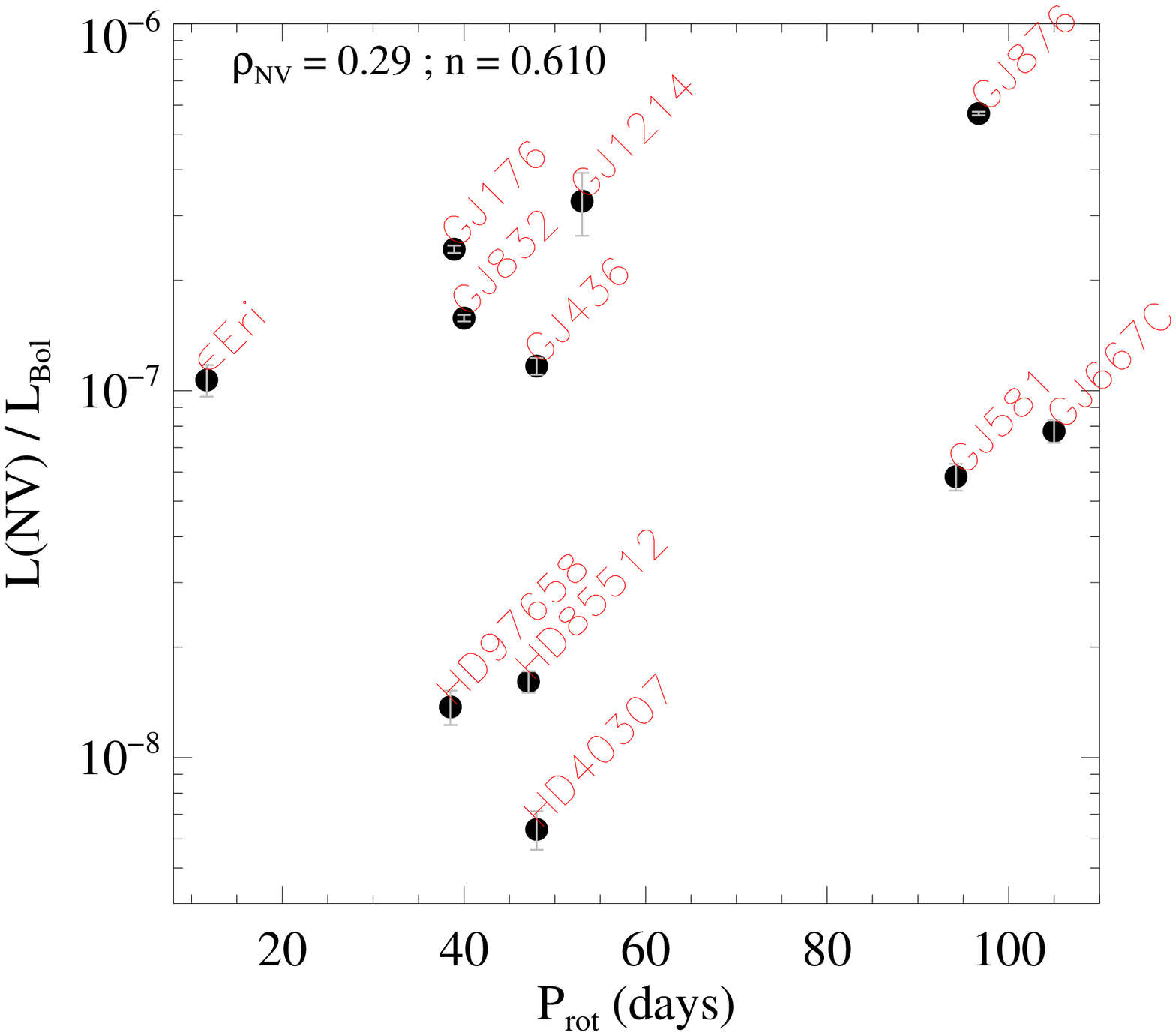,width=2.52in,angle=90}
\vspace{0.1in}
\caption{
\label{cosovly} ($top$) The transition region activity, traced here by the fractional \ion{N}{5} luminosity (relative to the bolometric luminosity), displays a weak anti-correlation with stellar effective temperature.  Note that $\epsilon$ Eri is the only ``active'' star in the MUSCLES sample. The middle plot shows that the total \ion{N}{5} luminosity may decline with rotation period, with the exception of GJ 876 (which is influenced by strong flares).  ($bottom$) There is no correlation between the fractional hot gas luminosity and the stellar rotation period.   }
\end{center}
\end{figure}

\begin{figure*}
\figurenum{12}
\begin{center} 
\epsfig{figure=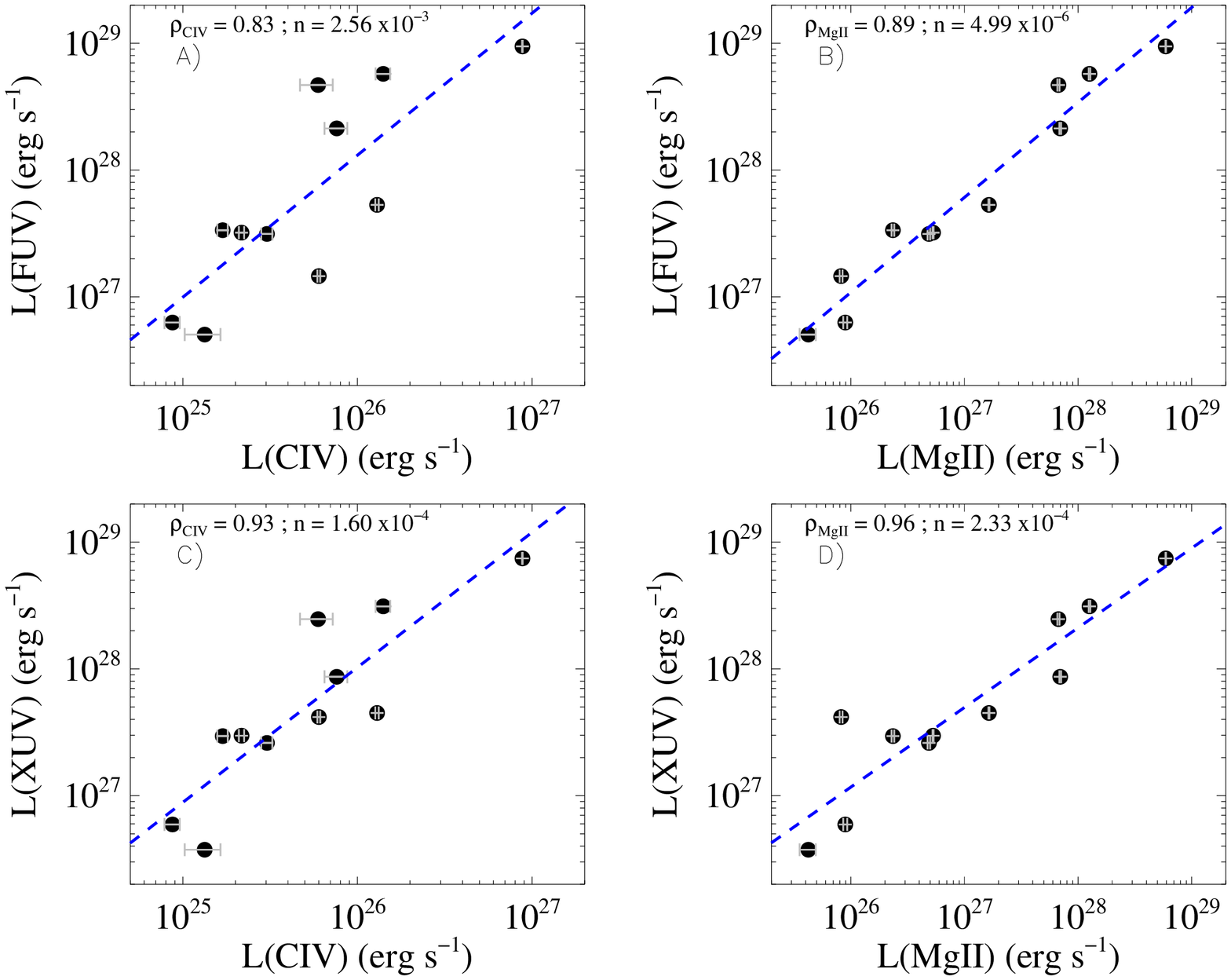,width=5.4in,angle=90}
\vspace{+0.16in}
\caption{
\label{cosovly} We demonstrate that both \ion{C}{4} and \ion{Mg}{2} are well-correlated with the total FUV and EUV fluxes.  A) shows the \ion{C}{4}-FUV luminosity relation and corresponding Pearson coefficient and statistical likelihood of a null correlation (top of each panel), B) shows the \ion{Mg}{2}-FUV luminosity relation, C) shows the \ion{C}{4}-XUV luminosity relation, and D) shows the \ion{Mg}{2}-XUV luminosity relation.  }
\end{center}
\end{figure*}

\subsection{The Correlation Between \ion{C}{4} and \ion{Mg}{2} Fluxes and Broadband Luminosities }

Another goal of the MUSCLES Treasury Survey is to identify individual spectral tracers that can serve as proxies for the broadband fluxes from low-mass stars so that large, resource-intensive projects may not be required to obtain accurate estimates for the energetic radiation environments around exoplanets in the future.   This anticipates a time beyond the current suite of UV and X-ray observatories (e.g., $XMM$, $Chandra$, and $HST$) capable of making these measurements, and a time when the sheer number of potentially habitable planets around low-mass stars precludes a detailed panchromatic characterization of every target (possibly after the $TESS$ mission).   In this case, simple relationships for acquiring reasonably accurate broadband stellar flux measurements will be essential for accurately modeling the atmospheric spectra from these worlds when they are acquired by future flagship missions in the 2020s, 2030s, and 2040s.  Below, we describe the relationship between the XUV and FUV luminosities of the MUSCLES stars with the most prominent FUV and NUV emission lines, \ion{Mg}{2} and \ion{C}{4}\footnote{The stellar Ly$\alpha$ emission line is the brightest FUV line in M dwarf spectra~\citep{france13}, however the total emission line flux cannot be retrieved without significant modeling analysis~\citep{wood05,youngblood15}}.   In a future work, we will explore ground-based tracers of the UV radiation environment (e.g., \ion{Ca}{2} H \& K fluxes and equivalent widths from our contemporaneous ground-based observations; Youngblood et al. 2016a~--~in preparation).  

Figure 12 shows the relationship between the broadband luminosities $L$(XUV) and $L$(FUV) and the emission line luminosities $L$(\ion{C}{4}) and $L$(\ion{Mg}{2}).   The figures show a strong correlation between all of these quantities, with Pearson correlation coefficients of [0.83, 0.86, 0.87, 0.90] for the relationships between [FUV~--~\ion{C}{4}, FUV~--~\ion{Mg}{2}, XUV~--~\ion{C}{4}, XUV~--~\ion{Mg}{2})], respectively.  The probability of a non-correlation is $<$ 3~$\times$~10$^{-3}$ for all four curves.   We present quantitative log-log relations\footnote{log$_{10}$$L$(band)~=~$m$~$\times$~log$_{10}$$L$(line) + $b$} in Table 2.  The RMS scatter around the [FUV~--~\ion{C}{4}, FUV~--~\ion{Mg}{2}, XUV~--~\ion{C}{4}, XUV~--~\ion{Mg}{2})] fit, ($L$(band) - $L$(fit))/$L$(band), is [153\%, 39\%, 109\%, 57\%], respectively.  

We have selected \ion{C}{4} and \ion{Mg}{2} because they are the two most readily-observable emission lines in the FUV and NUV bandpasses.  \ion{Mg}{2} shows a tighter correlation, but as we begin to probe exoplanetary systems at greater distances from the Sun, this relationship will become compromised by the additional contribution from interstellar \ion{Mg}{2} absorption components with radial velocities coincident with the stellar radial velocities~\citep{redfield02} as well as the possibility of gas-rich circumstellar environments fueled by mass-loss from short period gaseous planets~\citep{haswell12,fossati15}.  By contrast, \ion{C}{4} is essentially free from interstellar extinction out to the edge of the Local Bubble where dust opacity begins to contribute appreciably.  There are rare exceptions where hot gas in the Local Bubble can contribute small amounts of \ion{C}{4} attenuation~\citep{welsh10}, but the effect is considerably less than for \ion{Mg}{2}.

\subsection{Temporally Resolved Spectra and Energetic Flares}

As discussed in the introduction, temporal variability of the energetic radiation is considerably higher for M dwarfs than for solar type stars~\citep{mitra05}, with active M stars showing disk-integrated flux increases of an order-of-magnitude or more during large flares~\citep{hawley91,hawley03,osten05}.  The MUSCLES pilot program showed that even inactive M dwarfs could show impulsive flare behavior in their UV light curves~\citep{france12a}.  The UV monitoring component of the MUSCLES Treasury Survey 
was designed to present a uniform database of flares from exoplanet host stars so the flare frequency-amplitude relations could be derived for these stars and the impact of impulsive events on the atmospheres or orbiting planets could be assessed.  For cases where the X-ray observations were scheduled simultaneously, the overlapping observatory coverage was planned for the HST 5-orbit monitoring campaigns.  

The intention of this section is not to give a thorough quantitative description of the flare catalog produced in the survey; that work will be presented in a follow-on paper by~\citet{loyd16}.  This section is intended to introduce the variability data and present one example of an optically quiet M dwarf host star that is among the most UV/X-ray active sources ever observed.     Figure 13 shows the light curves of GJ 876, recorded in four bright emission lines, \ion{C}{2} $\lambda$$\lambda$1334, 1335 ($T_{form}$~$\sim$~3~$\times$~10$^{4}$ K), \ion{Si}{3} $\lambda$1206 ($T_{form}$~$\sim$~4~$\times$~10$^{4}$ K), \ion{Si}{4} $\lambda$$\lambda$1394, 1403 ($T_{form}$~$\sim$~6~$\times$~10$^{4}$ K), and \ion{N}{5} $\lambda$$\lambda$1239, 1243 ($T_{form}$~$\sim$~2~$\times$~10$^{5}$ K).   The data are binned to a 30 second cadence and displayed relative to the start time of the first observation.  The individual exposures are labeled for reference.   Several flares are immediately apparent, the strongest being during the last two orbits of the monitoring campaign.   The COS detector background, measured at the same dispersion direction location as the \ion{Si}{4} lines but offset below the spectral trace in the cross-dispersion direction, is also shown (orange squares) to demonstrate the stability of the instrument during these measurements.

Figure 14 ($top$) shows a zoom on the brightest UV flare, binned to a 10 second cadence, occurring near $T_{exp}$~=~17,400s.  Each light curve is normalized to unity during a pre-flare window (17,100~--~17,250s) to enable a comparison of the relative flare responses of each line.  The flare/quiescent flux increase in the brightest line, \ion{Si}{3} is $\sim$~110.   \ion{C}{2} and \ion{Si}{4} show flux increases of order $\sim$~50, and \ion{N}{5} and the FUV continuum increase by factors of $\sim$~5.  This level of luminosity increase indicates the largest relative UV flare ever directly detected in a disk-integrated observation of a star, despite the relative inactivity of GJ 876 suggested by the \ion{Ca}{2} activity index.  One observes that the lightcurves are line-dependent, with intermediate temperature ions showing a larger relative flux increase while the higher temperature ion (\ion{N}{5}) does not respond as strongly but shows a decay time several times longer than \ion{C}{2}, \ion{Si}{3}, and \ion{Si}{4} (Figure 13).  Figure 14, ($bottom$) shows time-resolved line ratios in the GJ 876 lightcurves.  Several emission lines are ratioed to \ion{Si}{4} and normalized to their pre-flare line ratios, so their relative change is meaningful.  In this way, we can place constraints on the atmospheric temperature regime where most of the $observable$ energy is deposited (our observations do not contain information about energy deposited in cooler or optically thick atmospheric layers).  The \ion{Si}{3}/\ion{Si}{4} ratio increases during the flare, while the \ion{C}{2}/\ion{Si}{4} and \ion{N}{5}/\ion{Si}{4} ratios, which sample gas both hotter and colder than \ion{Si}{3} are depressed.  This argues that the flare energy distribution is peaked near the \ion{Si}{3} formation temperature, roughly (4~--~5)~$\times$~10$^{4}$ K (with both low and high energy tails), and evolves during the flare.

\begin{figure}
\figurenum{13}
\begin{center}
\epsfig{figure=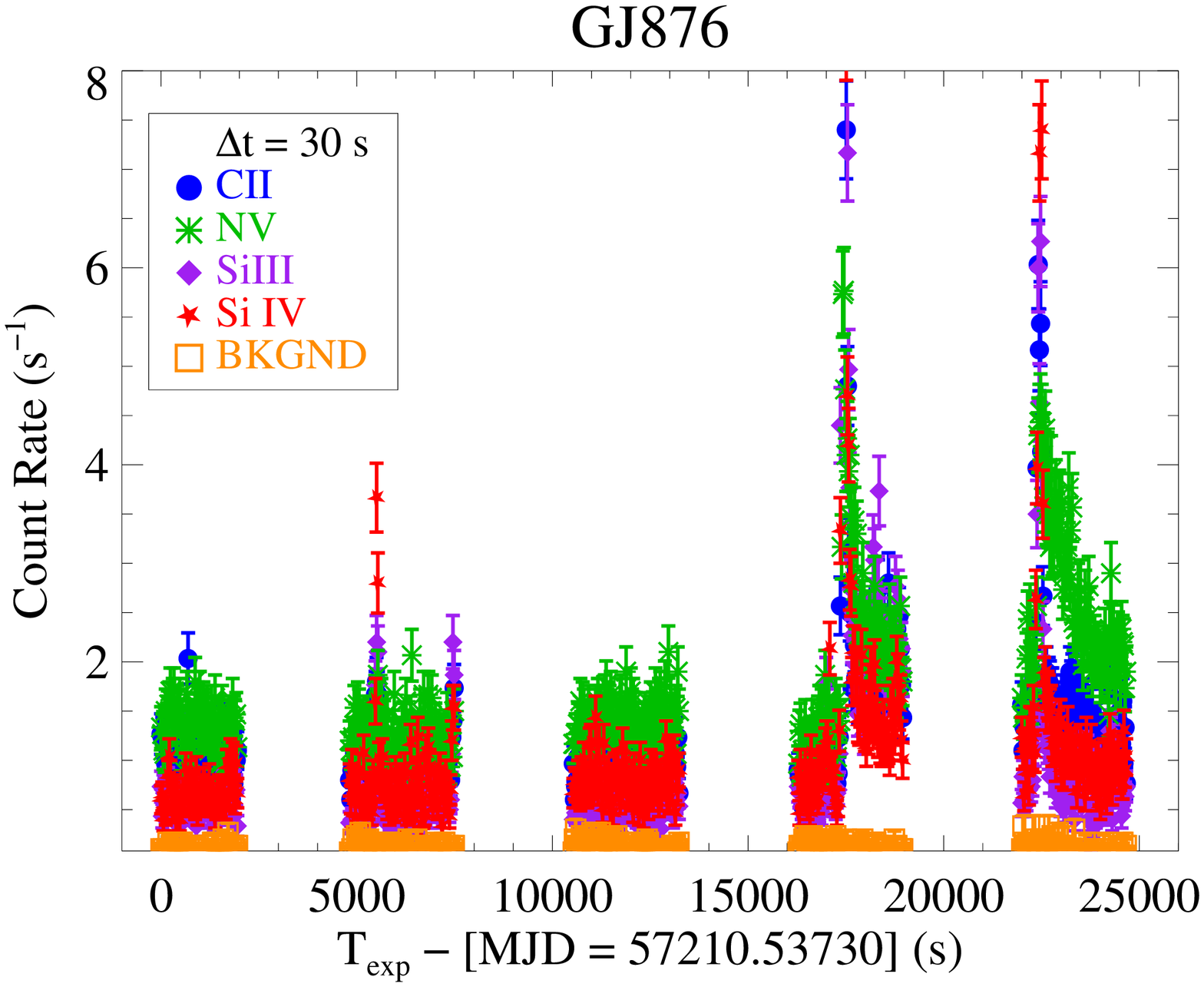,width=2.6in,angle=90}
\vspace{-0.1in}
\caption{\label{cosovly} An example of a flare lightcurve from the five contiguous $HST$ orbits with the COS G130M mode.  The G130M mode provides a wealth of temperature diagnostics from chromosphere and transition region, spanning $\approx$~3~$\times$~10$^{4}$ K (\ion{C}{2}) through $\approx$~2~$\times$~10$^{5}$ K (\ion{N}{5}).  The instrumental background is shown in orange.   The data are displayed at a 30 second cadence in this figure.
The strong flare observed during the fourth orbit is shown in detail in Figure 14.   }
\end{center}
\end{figure}

\begin{figure}
\figurenum{14}
\begin{center}
\epsfig{figure=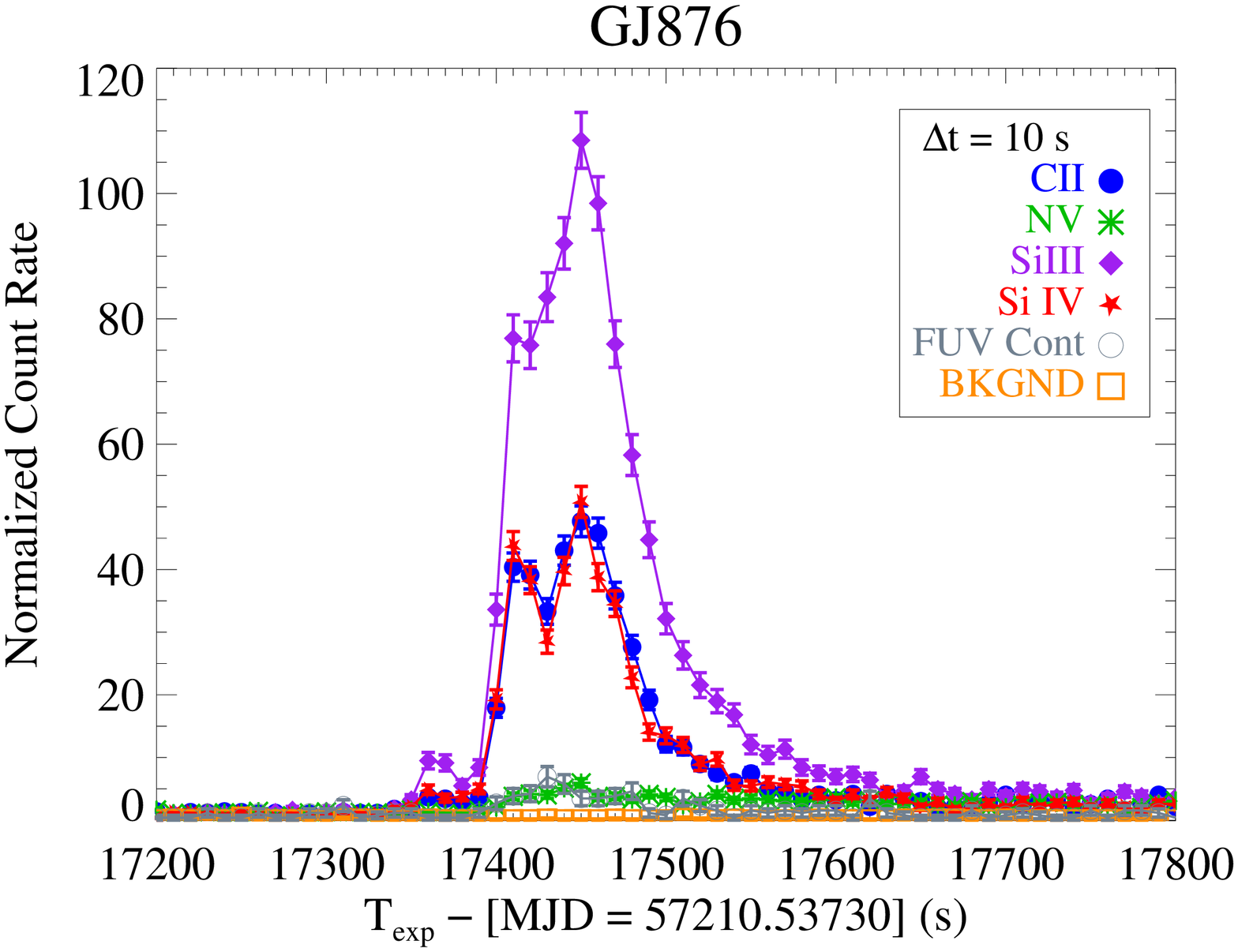,width=2.52in,angle=90}
\epsfig{figure=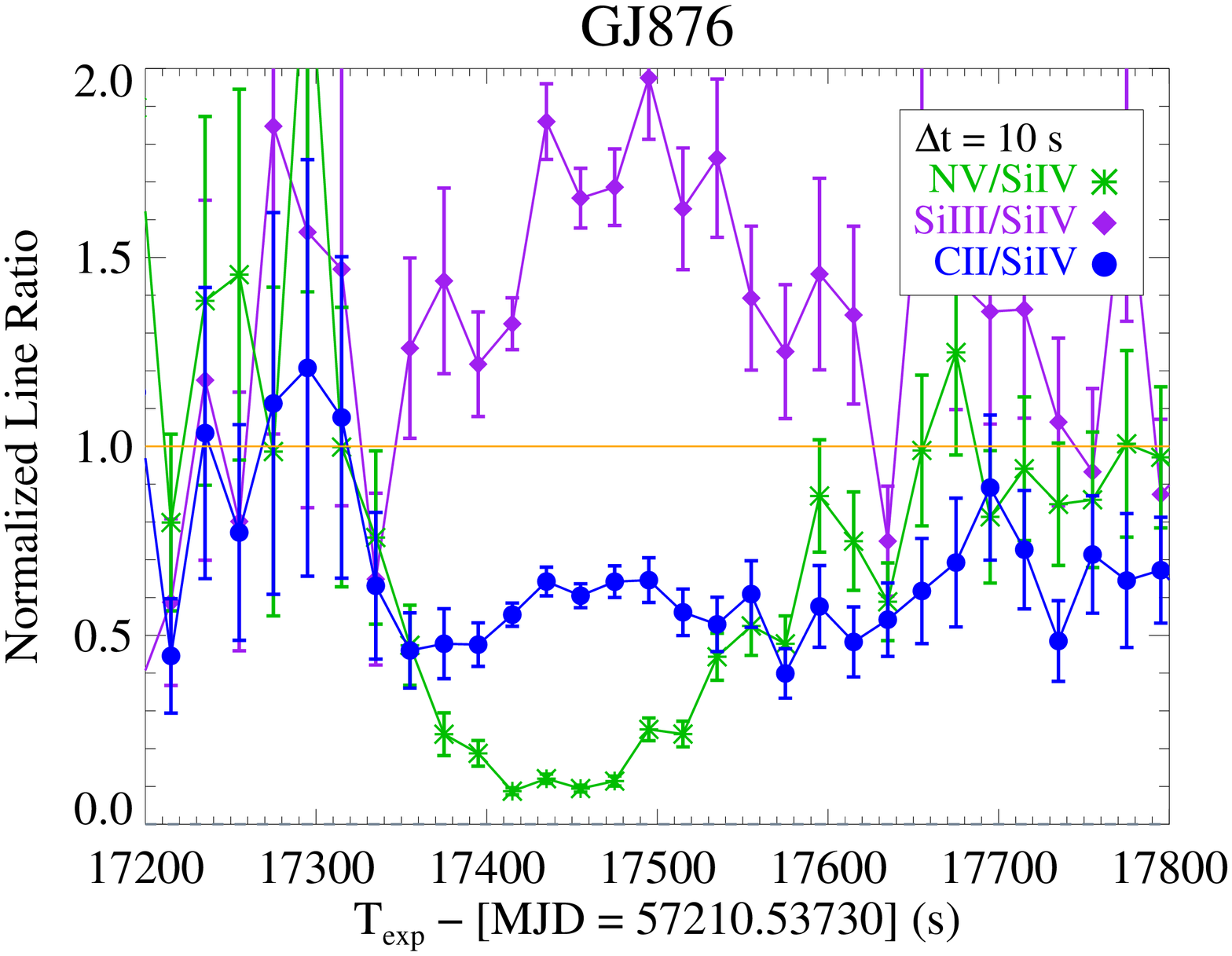,width=2.52in,angle=90}
\vspace{-0.1in}
\caption{
\label{cosovly} ($top$) A zoom-in on the strong UV flare observed on GJ 876 (Figure 13).  The peak flare/quiescent flux ratio for this flare in the Si$^{2+}$ ion is~$\approx$~110, making this one of the most extreme disk-integrated UV flares ever detected.   The bottom panel shows normalized emission line ratios for several diagnostic lines and their evolution during the flare.  This flare peaked in the \ion{Si}{3} ion, with  \ion{Si}{3}/\ion{Si}{4} ratios increasing during the flare.  Less energy was deposited in the hotter and cooler ions, represented here by \ion{N}{5}/\ion{Si}{4} and \ion{C}{2}/\ion{Si}{4} ratios, respectively.  The full MUSCLES flare catalog will be presented by Loyd et al. (2016 - in prep).   }
\end{center}
\end{figure}

\begin{figure}
\figurenum{15}
\begin{center}
\epsfig{figure=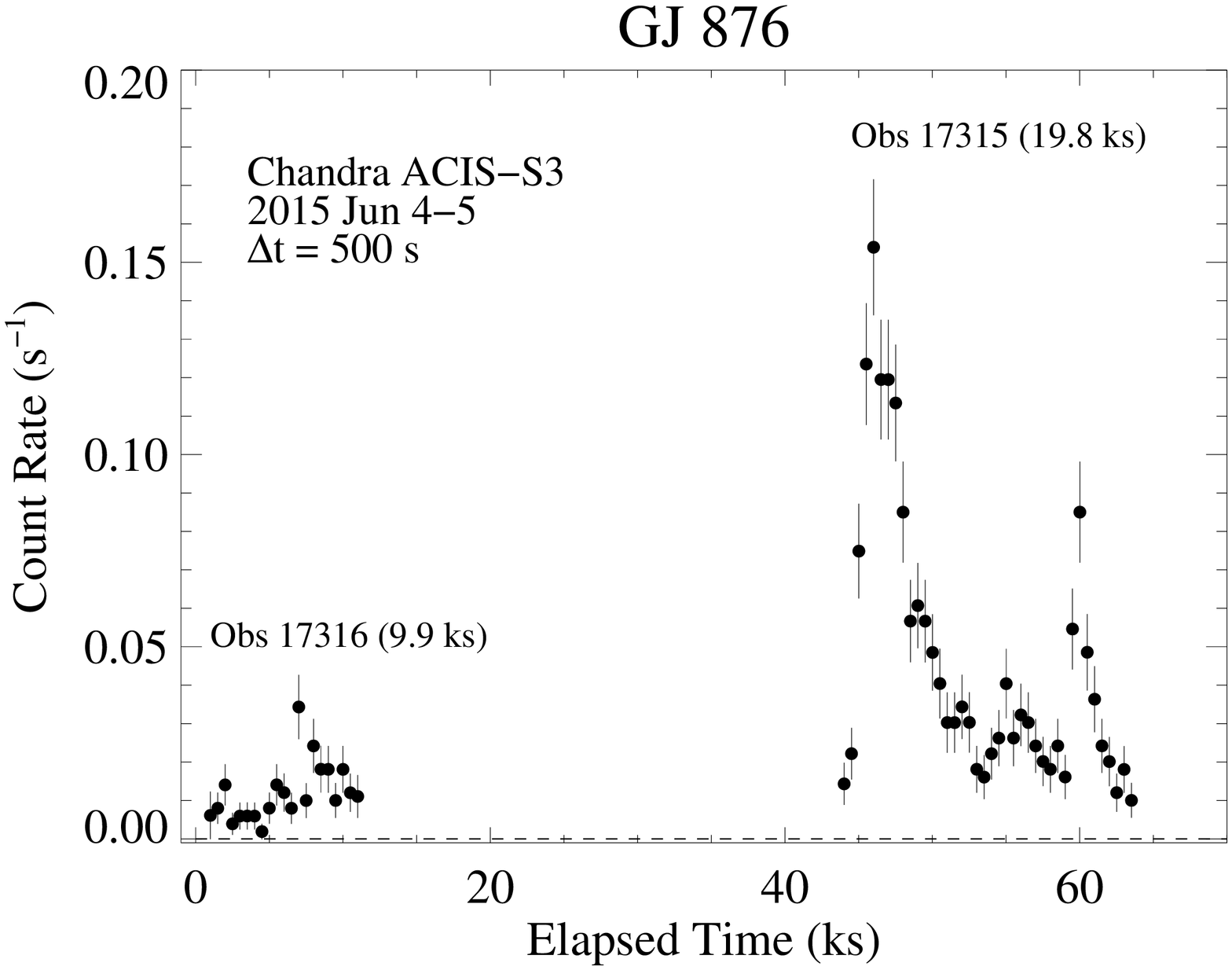,width=2.6in,angle=90}
\vspace{-0.1in}
\caption{
\label{cosovly} An example X-ray flare lightcurve of GJ 876 from the $Chandra$ program to support the five contiguous monitoring orbits carried out with COS G130M.  An $HST$ safing event prevented this lightcurve from being simultaneous with the FUV light curves shown in Figures 13 and 14. }
\end{center}
\end{figure}

In order to understand the impact of these flares on the planets in these systems, we need to convert the observed emission line flares to an estimate of the broadband UV flare energy.  Converting the raw spectral light curves into flux units by comparing the orbit-averaged count rates and stellar emission line fluxes, we can create flux-calibrated lightcurves~\citep{loyd14}.   The total flare energy in a given emission line is then 
\begin{equation}
E(\lambda)~=~4 \pi d^{2} \int_{t_{start}}^{t_{end}} F(\lambda) dt  
\end{equation}
where $F$($\lambda$) is the line flux in (erg cm$^{-2}$ s$^{-1}$), $d$ is the stellar distance, and $t_{start}$ and $t_{end}$ are the initial flare rise times and the time when the flare returns to the quiescent level, respectively.  For the bright flare considered here, the exposure ends before the ionic lines return to the quiescent levels, so we set $t_{end}$ as end of the orbit in this case.  The \ion{Si}{3} and \ion{Si}{4} emission lines are representative of the upper chromosphere and transition region emission formed between $\sim$~30~--~80 kK; we use these lines to estimate the total flare emission from these regions.  The total energies in these lines are log$_{10}$ $E$(\ion{Si}{3}) = 29.17 erg and log$_{10}$ $E$(\ion{Si}{4}) = 29.08 erg.  We use the M dwarf stellar atmosphere model of~\citet{fontenla15} to estimate the total emission in the 300~--~1700~\AA\ range that contains the majority of transition region emission.  Emission at $\lambda$~$<$~300~\AA\ is mainly coronal in origin while emission at $\lambda$~$>$~1700~\AA\ is mainly cooler chromospheric gas.  

The ionized silicon flux energies can be converted to a broadband XUV~+~FUV (300~--~1700\AA) energy by computing the fractional flux emitted in these lines and the fractional radiative cooling rate that is contributed by 30~--~80 kK gas.     The total broadband energy is then

\begin{equation}
    \begin{split}
E(300 - 1700 \AA)~=~E(Si III + Si IV) \\ 
	\times~( f_{300 - 900} + f_{900 - 1200} + f_{1200 - 1700} )  \\ 
	 \times C_{30-80kK}
   \end{split}	 
\end{equation}
where $f_{300 - 900}$ is the ratio of stellar flux in the 300~--~900\AA\ band to the combined \ion{Si}{3} + \ion{Si}{4} flux, $F$(300~--~900\AA)/$F$(\ion{Si}{3} + \ion{Si}{4}).  Similarly,   $f_{900 - 1200}$~=~$F$(900~--~1200\AA)/$F$(\ion{Si}{3} + \ion{Si}{4}) and $f_{1200 - 1700}$~=~$F$(1200~--~1700\AA)/$F$(\ion{Si}{3} + \ion{Si}{4}).  The $f_{300 - 900}$, $f_{900 - 1200}$, and $f_{1200 - 1700}$ 
ratios from the model atmosphere of~\citet{fontenla15} are 32, 14, and 290, respectively.  $C_{30-80kK}$ is the fraction of the total radiative cooling rate from the upper stellar atmosphere (6000 K~--~10$^{6}$ K) contributed by 30~--~80 kK gas, $C_{30-80kK}$~=~$\Gamma$(30~--~80 kK)/$\Gamma$(6000 K~--~10$^{6}$ K), where $\Gamma$ is the radiative cooling rate in units of  erg cm$^{-3}$ s$^{-1}$.  Using the cooling rates for this model atmosphere~\citep{fontenla15}, we compute $C_{30-80kK}$~=~16.8\%.  Note that these are the equilibrium cooling rates for the quiescent model atmosphere, and may be different during the post-reconnection heating associated with the flare.  Combining these elements, we estimate that the total UV flare energy associated with this event is log$_{10}$ $E$(300~--~1700\AA) = 31.18 erg, comparable to the total quiescent luminosity of the star $E_{flare}$(UV)~$\sim$~0.3 $L_{*}$$\Delta$$t$ ($\Delta$$t$ = 1 second).   


In a future work,~\citet{loyd16} will show that $HST$ monitoring observations are able to measure and quantify the M dwarf flare amplitude--frequency distribution.  Critically, $HST$ provides the sensitivity and spectral resolution to analyze light curves covering a range of formation temperatures, and capture the short duration events that dominate the flare distribution.  A dedicated $HST$ spectroscopic flare monitoring program is currently the best avenue for understanding the energy and temporal distribution of flares and their potential influence on low-mass planets


\subsubsection{ X-ray Lightcurves}
When possible, X-ray observations were taken in concert with the UV light curves described here.  However, this was only possible for about half of the MUSCLES observations.  Figure 15 shows the non-simultaneous X-ray light curve of GJ 876\footnote{This observations was scheduled simultaneously, but $HST$ entered safe mode during the scheduled period during early June 2015 and the $Chandra$ observations executed without $HST$.  The $HST$ observations were rescheduled about a month later, without supporting X-ray observations.}, finding another large flare on this object.  The X-ray spectra of both the quiescent and flare periods have been fitted, and the overall X-ray luminosity increased by a factor of $\sim$~10 during the flare.  The actual luminosity increase was likely larger but diluted by the cadence adopted to provide sufficient S/N in each time bin.  Given the similarity of the UV and X-ray light curves of dMe stars observed by $XMM$~\citep{mitra05}, it seems likely that these large flares are similar events and occur with regularity on GJ 876.  We note that~\citet{poppenhaeger10} also noted a high level of X-ray activity on GJ 876 and~\citet{france12a} observed a large UV flare in the very first MUSCLES observations.   This apparently inactive star displays a flare outburst almost every time it is observed at wavelengths below the atmospheric cut-off.  While we have elected to focus on the most UV active star in the MUSCLES data set, roughly half of our targets displayed UV flare activity.  This will be explored in greater detail in~\citet{loyd16}.

\section{Discussion}



\subsection{Chromospheric and Transition Region Activity of GJ 1214: On, Off, or Variable?}

We use GJ 1214, the only star in our sample with effective temperature $<$~3000~K, to investigate the evolution of upper atmosphere activity towards the stellar/sub-stellar boundary.  In the MUSCLES pilot study, we presented an upper limit to the Ly$\alpha$ flux of GJ 1214 based on earlier STIS G140M observations.  GJ 1214 has a spectral type of M4.5V, with $T_{eff}$~=~2949~$\pm$~30 K (Kundurthy et al. 2011; consistent with the 2935 K derived from the photospheric model fitting described in Paper III).  It was the most distant source in the MUSCLES pilot study, and the only star for which \ion{H}{1} Ly$\alpha$ was not detected.    This non-detection was  approximately a factor of 10 below the expected flux level based on an extrapolation of the F(Ly$\alpha$) versus F(\ion{Mg}{2}) relation~\citep{france13} and was made more surprising by the solid detection of \ion{C}{4} $\lambda$~1548 emission.   This led us to speculate about the possibility of a high molecular fraction atmosphere that suppresses atomic hydrogen emission or possibly an `on/off' behavior where the basal flux level is very low and the chromospheric and coronal emission can only be observed during flares (as has been noted for the M8 star VB 10 by Linsky et al. 1995, however more recent observations demonstrated an `always on' behavior; Hawley \& Johns-Krull 2003).\nocite{linsky95,hawley03b} Subsequently, X-rays from GJ 1214 were detected by {\it XMM-Newton}~\citep{lalitha14}.  All of the source counts during the $XMM$ observations occurred in a single time bin, suggesting that the `on/off' scenario may be correct.  

The new MUSCLES observations instead argue for a weak but persistent high-energy flux from GJ 1214.   We have detected the full complement of FUV and NUV emission lines from GJ 1214 with the deeper MUSCLES treasury data (including direct observation of Ly$\alpha$, see Paper II; Table 3).  While the Ly$\alpha$ reconstruction is uncertain owing to the low-S/N and the small fraction of the line profile that is detected, we can compare the observed flux from the star over multiple epochs.  The Ly$\alpha$ (STIS G140M) and \ion{C}{4} + \ion{Mg}{2} (COS G160M + STIS G230L) observations in~\citet{france13} were separated by roughly 15 months, while the 2015 MUSCLES Treasury data were obtained over a period of $<$ 2 days.   We find that the observed flux level from GJ 1214 in 2015, $F_{2015}$(Ly$\alpha$) = 1.8 ($\pm$ 0.3)~$\times$~10$^{-15}$ erg cm$^{-2}$ s$^{-1}$, is consistent with the upper limit presented previously,  $F_{2013}$(Ly$\alpha$) $\leq$ 2.4 ~$\times$~10$^{-15}$ erg cm$^{-2}$ s$^{-1}$.  The reconstructed flux is approximately twice this value $F_{2015,recon}$(Ly$\alpha$) = 5.5~$\times$~10$^{-15}$ erg cm$^{-2}$ s$^{-1}$, which is still roughly a factor of three lower than the expected reconstructed Ly$\alpha$ flux based on the $F$(Ly$\alpha$) versus $F$(\ion{Mg}{2}) relation~\citep{youngblood15}.   We find that the total \ion{C}{4} brightness has increased by a factor of 2 compared to our earlier study ($F_{2015}$(\ion{C}{4}) = 5.2 ($\pm$ 1.2)~$\times$~10$^{-16}$ erg cm$^{-2}$ s$^{-1}$ versus $F_{2013}$(\ion{C}{4}) = 2.6 ($\pm$ 0.5)~$\times$~10$^{-16}$ erg cm$^{-2}$ s$^{-1}$), while the \ion{Mg}{2} flux is approximately constant ($F_{2015}$(\ion{Mg}{2}) = 1.7 ($\pm$ 0.3)~$\times$~10$^{-15}$ erg cm$^{-2}$ s$^{-1}$ versus $F_{2013}$(\ion{Mg}{2}) = 2.2 ($\pm$ 0.2)~$\times$~10$^{-16}$ erg cm$^{-2}$ s$^{-1}$).  

GJ 1214 displays fractional luminosities that are typical for the mid/late-M dwarfs in the MUSCLES survey (Table 3; Figures 10 and 11).  It has the lowest absolute levels of high-energy radiation (Figure 8) owing to its low mass.  The low intrinsic luminosity level and the star's relatively large distance ($d$~=~13 pc) can explain the non-detection of Ly$\alpha$ in the previous work.   It appears that GJ 1214 is simply a scaled-down version of the typical planet hosting mid-M dwarf, and not in a fundamentally different state of chromospheric activity.  While GJ 1214 clearly has flares, we have now shown that the basal flux level of this star is similar to other low-mass exoplanet host stars.  Combining this result with the high activity levels on M5 stars such as GJ 876, we conclude that special time-dependent photochemistry (other than the incorporation of impulsive flares and longer term variability; Section 1) is not necessary down to spectral type M5.

\subsection{Comparison of UV Observations to Coronal Models}  

The MUSCLES panchromatic SED creation relies on direct observation at all wavelengths except at red/IR wavelengths ($\lambda$~$>$~6000~\AA) and the XUV/FUV region between 50~--~1170~\AA.  As discussed in the introduction, the EUV regulates heating and mass-loss in planetary atmosphere, particularly for short-period planets around stars with large EUV and X-ray fluxes.  However, obtaining a complete EUV spectrum of any cool star other than the Sun is currently impossible owing to attenuation by the ISM.  Our approach to filling in this observationally inaccessible gap is to employ a coronal model from 50~--~100~\AA\ (a single or two-temperature APEC model, Smith et al. 2001) and models of solar active regions from 100~--~1170~\AA~\citep{fontenla11,linsky14}.   This approach assumes a scaling between the chromospheric Ly$\alpha$ emission and the chromospheric + transition region + coronal flux that contributes to the EUV bands, particularly between 100~--~400~\AA.\nocite{smith01}  

An alternative approach has been taken by~\citet{forcada11}, who used X-ray observations of the coronae of exoplanet host stars and an emission measure distribution technique to predict the EUV and part of the FUV spectra of low-mass host stars.  The synthetic spectral output of coronal models for three of the MUSCLES stars ($\epsilon$ Eri, GJ 436, and GJ 876) are available on the {\tt X-exoplanets} website\footnote{\tt http://sdc.cab.inta-csic.es/xexoplanets/jsp/homepage.jsp}, and provide overlap with the high S/N MUSCLES data in the 1150~--~1200~\AA\ region.   We downloaded these synthetic spectra to compare the coronal model fluxes with the data for the bright \ion{C}{3} $\lambda$1175 multiplet.  These lines are formed in the transition region at T$_{form}$~$\sim$~6~$\times$~10$^{4}$~K, and therefore provide a good diagnostic for how well coronal models are able to reproduce the UV emission throughout the FUV.   The {\tt X-exoplanets} synthetic spectra are provided in units of photons s$^{-1}$ cm$^{-2}$ bin$^{-1}$, with a variable bin size. 
The spectra are multiplied by a factor of (bin size)$^{-1}$ and then integrated over wavelength to compute integrated \ion{C}{3} line photons s$^{-1}$ cm$^{-2}$.  The MUSCLES data were converted to photons and integrated over the same wavelength interval (1174~--~1177~\AA).   

We find that the coronal models systematically underpredict the UV emission line strengths by factors of a few to tens (Table 4).  For $\epsilon$ Eri, we find $F_{data}$(\ion{C}{3})~=~1.8~$\times$~10$^{-2}$~photons s$^{-1}$ cm$^{-2}$ versus $F_{Xexoplanets}$(\ion{C}{3})~=~6.6~$\times$~10$^{-4}$~photons s$^{-1}$ cm$^{-2}$.  For GJ 436, we find $F_{data}$(\ion{C}{3})~=~4.2~$\times$~10$^{-5}$~photons s$^{-1}$ cm$^{-2}$ versus $F_{Xexoplanets}$(\ion{C}{3})~=~8.0~$\times$~10$^{-6}$~photons s$^{-1}$ cm$^{-2}$.  For GJ 876, we find $F_{data}$(\ion{C}{3})~=~6.5~$\times$~10$^{-4}$~photons s$^{-1}$ cm$^{-2}$ versus $F_{Xexoplanets}$(\ion{C}{3})~=~1.9~$\times$~10$^{-5}$~photons s$^{-1}$ cm$^{-2}$.  The coronal models underpredict these transition region fluxes by factors of approximately 27, 5, and 33 for $\epsilon$~Eri, GJ 436, and GJ 876, respectively.   This difference underscores the importance of having empirical inputs for transition region and chromospheric emission in the calculations of the EUV flux from cool stars (see also the discussion presented in Linsky et al. 2014). The systematic underprediction of the transition region emission brings into question the accuracy of the coronal models in the 400~--~900~\AA\ EUV region that is dominated by chromospheric and transition region line and continuum spectra.  Another important factor is the time variability in the coronal and chromospheric emission from these stars, which may not be the same.  The MUSCLES database provides an excellent resource for emission measure distribution-based stellar atmosphere models that simultaneously take into account both the intermediate and high-temperature regions of the stellar atmosphere (e.g., Chadney et al. 2015) and may provide more accurate EUV flux estimates for the exoplanet community.\nocite{chadney15}

\subsection{Star-Planet Interactions Observed in Transition Region Emission Lines}

It has been suggested that stellar and exoplanetary magnetic fields can interact in exoplanetary systems~\citep{shkolnik03,lanza08}, possibly manifesting as enhanced stellar activity relative to nominal age-rotation-activity relationships for isolated main sequence stars~\citep{barnes07,mamajek08}.  The magnitude of this interaction, as measured by the energy dissipated in the stellar atmosphere, may be proportional to the strength of the stellar magnetic field, the planetary magnetic field, and the relative speed of the planet's orbital velocity compared to the stellar magnetic rotation rate.  While the stellar magnetic field and orbital velocity can be readily measured, exoplanet magnetic fields have proven notoriously hard to detect (see, e.g., Lecavelier des Etangs 2011 and Hallinan et al. 2013), with very few possible detections from low-frequency radio observations ~\citep{lecavelier13} and early-ingress measurements of NUV and optical transit light curves~\citep{fossati10,lai10,vidotto10,llama11,cauley15}.\nocite{lecavelier11,hallinan13,shkolnik05,kashyap08}  If the above dependencies on the magnetic field and velocities are valid, then a general trend of enhanced energy dissipation, which could be observed as enhanced emission from the stellar corona (X-rays; Kashyap et al. 2008) or chromosphere (\ion{Ca}{2}; Shkolnik et al. 2005), should correlate with $M_{plan}$/$a_{plan}$, where $M_{plan}$ is the planetary mass and $a_{plan}$ is the semi-major axis.  Because magnetic field strength increases with planetary mass in the solar system, one would expect that the most massive, closest-in planets in exoplanetary systems should produce the largest signal on their host stars.  However, almost every claimed detection of a modulation in activity on exoplanet host stars or connection with orbital phase has not been confirmed by follow up observation or re-analysis of the same data set (e.g., Shkolnik et al. 2008; Poppenhaeger et al. 2010).

\begin{figure}
\figurenum{16}
\begin{center}
\epsfig{figure=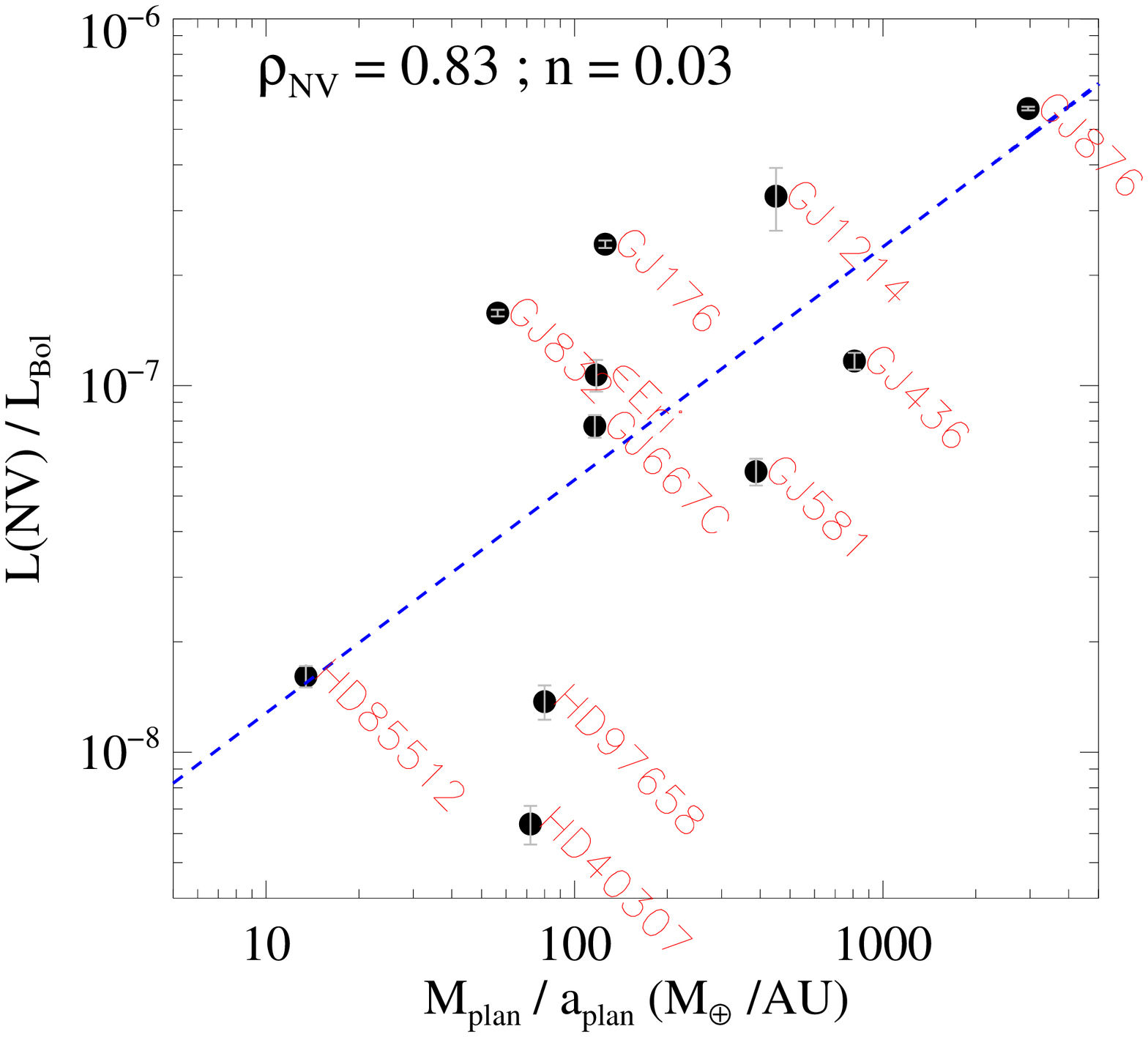,width=2.6in,angle=90}
\vspace{+0.1in}
\caption{
\label{cosovly} The fractional \ion{N}{5} luminosity is positively correlated with a measure of the `star-planet interaction strength', $M_{plan}$/$a_{plan}$~\citep{shkolnik13}, where $M_{plan}$ is the mass of the most massive planet in the system (in Earth masses) and $a_{plan}$ is the semi-major axis of the most massive planet in the system (in AU).  The Pearson coefficient and statistical likelihood of a null correlation is shown at top.  This provides tentative evidence that the presence of short-period planets enhances the transition region activity on low-mass stars, possibly through the interaction of their magnetospheres.  }
\end{center}
\end{figure}

\begin{figure}
\figurenum{17}
\begin{center}
\epsfig{figure=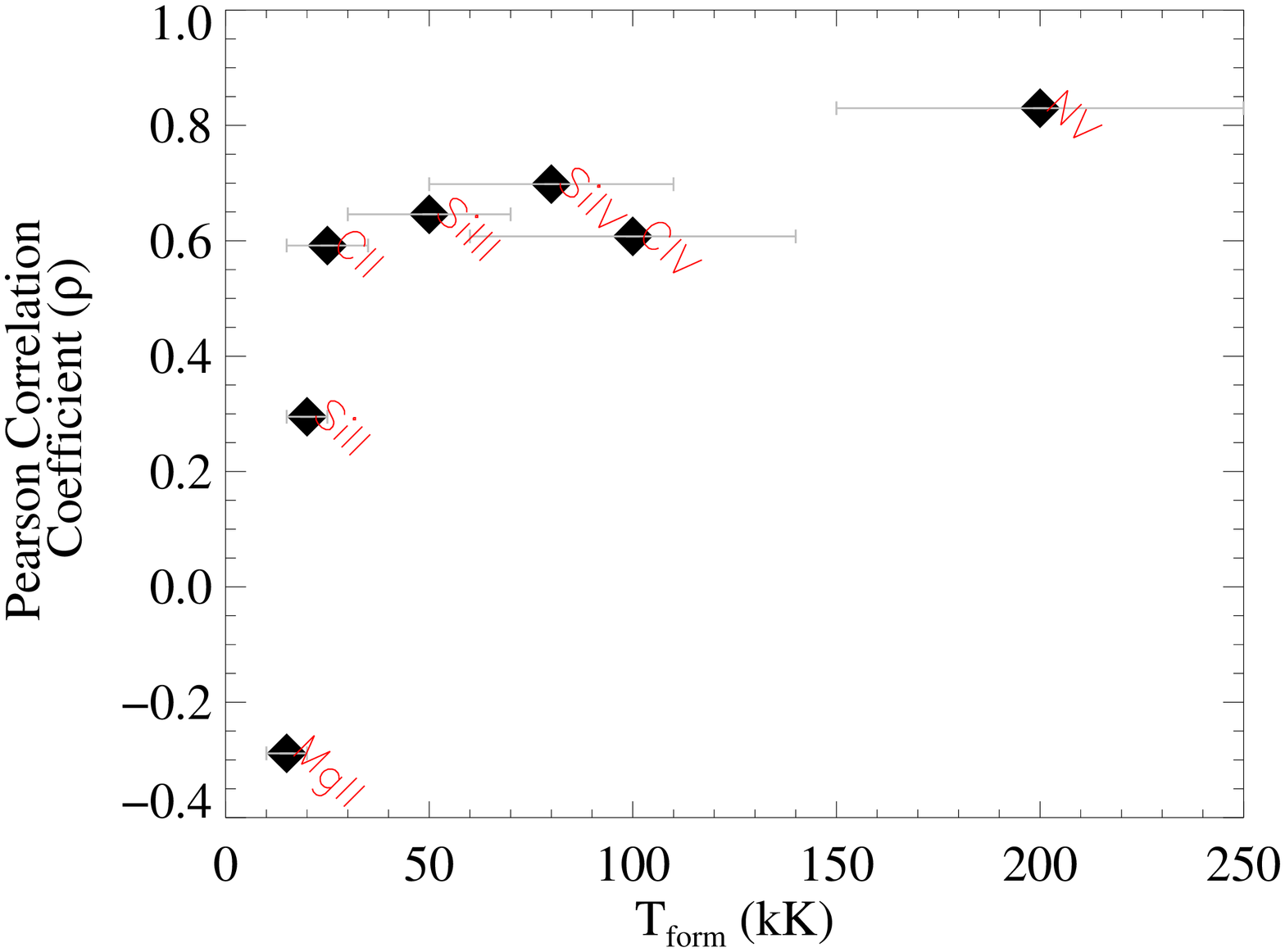,width=2.6in,angle=90}
\vspace{-0.0in}
\caption{
\label{cosovly} The Pearson correlation coefficients characterizing the strength of the relationship between the fractional emission luminosity in strong chromospheric and transition region lines and $M_{plan}$/$a_{plan}$.  These are plotted as a function of formation temperature to better understand where the SPI process is depositing energy into the stellar atmosphere.  Specific transitions are labeled in red, and the range of formation temperatures~\citep{fontenla15} are shown as the gray error bars.  We observe no significant correlation for formation temperatures $<$~5~$\times$~10$^{4}$~K, while the highest formation temperatures show the strongest correlations.  This suggests that the SPI deposits significantly more energy in the transition region than in the chromosphere for M stars.   }
\end{center}
\end{figure}

The search for star-planet interaction (SPI) has predominantly focused on hot Jupiters orbiting F, G, and K stars.  The MUSCLES database allows us to examine potential star-planet interactions for a range of planetary masses and semi-major axes on a new class of optically inactive low-mass star.  Furthermore, the MUSCLES database allows us to explore SPI as a function of emission line formation temperature, which can constrain the possible location of magnetic field line connection in the stellar atmosphere and subsequent location of the plasma heating.   Figure 16 shows a correlation between the \ion{N}{5} fractional luminosity and $M_{plan}$/$a_{plan}$.   The \ion{N}{5}~--~$M_{plan}$/$a_{plan}$ relation has a Pearson correlation coefficient of 0.83 and a statistical non-correlation likelihood of 0.035.   This suggests that the systems with close-in, massive planets may indeed be generating enhanced transition region activity, as probed by this $\sim$~2~$\times$~10$^{5}$ K gas\footnote{For systems with additional, lower-mass planets, we also computed the correlation between $L$(line)/$L_{Bol}$ and the summed SPI integration strength for all $j$ planets in the system, $\Sigma_{j}$~($M_{j}$/$a_{j}$), and find that the correlation is somewhat weakened, Pearson coefficient = 0.82 and $n$~=~0.2.}.   On the other hand, we do not observe a correlation with the lower-temperature chromospheric gas traced by \ion{Mg}{2} ($T_{form}$~$\sim$~(1~--~2)~$\times$~10$^{4}$ K).  The \ion{Mg}{2}~--~$M_{plan}$/$a_{plan}$ relation, for instance, displays a correlation coefficient and probability of a null-correlation, $-$0.3 and 0.06, respectively.   Figure 17 presents the Pearson correlation coefficients for the relation between the fractional luminosity and $M_{plan}$/$a_{plan}$ for several lines in the MUSCLES spectra.  While the null-hypothesis cannot be ruled out with high confidence for individual hot gas lines (i.e., the $n$ values are not very low), the trend with line formation temperature suggests that the connection between hot gas emission enhancement and $M_{plan}$/$a_{plan}$ may be real.    This suggests that the SPI correlation would not be observed in \ion{Ca}{2}, which has a comparable formation temperature distribution as \ion{Mg}{2} in M dwarfs~\citep{fontenla15}.    

 The correlation of the hot gas lines in the MUSCLES sample is somewhat surprising given the null result found by~\citet{shkolnik13} for $GALEX$ fluxes.  As the $GALEX$ observations are broader band measurements, there is an uncertain and variable contribution from the stellar photosphere that must be subtracted, lines from a range of formation temperatures are simultaneously included, and the highest temperature lines (e.g., \ion{N}{5}) are excluded.  Therefore, potential SPI signals are significantly diluted in UV imaging surveys relative to the MUSCLES survey, which focuses on lower mass stars with less continuum emission and spectrally resolved line-profiles that can be robustly measured.  This result argues that very low spectral resolution ($R$~$<$~500) observations at FUV or NUV wavelengths are not a promising avenue for discovering SPI-driven activity enhancements in low-mass stars.  
 
 We emphasize that given the small sample size, the detection of SPI in the MUSCLES sample should viewed with caution.  One explanation for these results is a simple scaling with stellar effective temperature.   Our inactive K dwarfs have low fractional luminosities and low-mass planetary systems, driving one end of the correlations in some cases.  A larger sample that is better controlled for systematics is needed to confirm the results presented here.  A future spectroscopic survey of G, K, and M exoplanet host stars with the COS G130M mode (to cover \ion{N}{5}, \ion{Si}{4}, and \ion{C}{2}) could be a promising data set with which to characterize enhanced magnetospheric stellar emission generated by SPI.

\section{Summary}

We have presented the first panchromatic survey of M and K dwarf exoplanet host stars from X-ray to UV to optical to IR wavelengths.  The MUSCLES Treasury Survey is built upon contemporaneous $Chandra$ (or $XMM$), $HST$, and ground-based data.  The 5~\AA\ to 5~$\mu$m SEDs have been assembled and hosted as high-level science products on the MAST website.  The main purpose of this paper was to present the motivation and overview of the MUSCLES Treasury Survey; more detailed analyses of the spectrally and temporally resolved SEDs, their impact of atmospheric photochemistry, and the suitability of M dwarfs as habitable planet hosts will be addressed by our team in future publications.    

  The main results of this work are: 
\begin{enumerate}
	\item All stars show energetic radiation (X-ray through UV) at all times during the observations.  Chromospheric, transition region, and coronal emission is directly observed from all targets.  Despite all but one of our targets having H$\alpha$ in absorption (that is, ``inactive'' in the traditional optical sense), all of the MUSCLES stars are X-ray and UV active.

	\item We have provided empirically-derived relations to compute the FUV and XUV stellar luminosity from \ion{Mg}{2} and \ion{C}{4} emission line fluxes.   

	\item The FUV/NUV flux ratio, an indicator for the potential abiotic formation of O$_{2}$ and O$_{3}$, declines with increasing stellar effective temperature by more than two orders of magnitude from $T_{eff}$~=~3000~--~5000 K.   Consequently, the FUV/NUV flux ratio declines by more than two orders of magnitude as habitable zone orbital distances increase from 0.1~--~0.7 AU.  The total FUV radiation field strength increases by factors of 2~--~3 over this distance, while the XUV radiation field strength is approximately constant.  The average FUV and XUV fluxes in the habitable zones of all K and M dwarfs studied are~$\approx$~10~--~70 erg cm$^{-2}$ s$^{-1}$.   {\it The spectral energy distribution of the radiation field changes dramatically for different habitable zone distances around low-mass stars, but the intensity of the high-energy radiation field does not.}  

	\item Despite their weak optical activity indicators (e.g., \ion{Ca}{2} emission core equivalent widths), several of our stars display extremely strong UV and X-ray flares. Flare/quiescent flare increases by a factor of $\sim$~10 are common on at least half of our stars with the strongest flares showing $E_{flare}$(UV)~$\sim$~0.3 $L_{*}$$\Delta$$t$.  

	\item Emission measure distribution models based on X-ray (coronal) data alone underestimate the FUV transition region flux by factors of~$\sim$~5~--~30, meaning that these models should not be used for calculating the FUV radiation field of exoplanet host stars.   The MUSCLES database provides an excellent resource for emission measure modeling of the stellar atmosphere where lines formed in the chromosphere, transition region, and corona can all be taken into account.   

	\item We present tentative evidence for star-planet interaction by measuring the fractional emission line luminosity as a function of the star-planet interaction strength, $M_{plan}$/$a_{plan}$.   Only the high-temperature transition region lines (\ion{C}{4} and \ion{N}{5}) show a positive correlation.  No correlation exists for lines with formation temperatures below~10$^{5}$~K, suggesting the interaction takes place primarily in the transition region or corona.   Moderate resolution FUV spectroscopy appears to be a promising avenue to further characterize star-planet interaction, while narrow-band FUV and NUV imaging are of less utility for characterizing the interactions of stars and planets owing to bandpass and spectral resolution limitations.  

\end{enumerate}

\acknowledgments
The data presented here were obtained as part of the $HST$ Guest Observing programs \#12464 and \#13650 as well as the COS Science Team Guaranteed Time programs \#12034 and \#12035.  This work was supported by STScI grants HST-GO-12464.01 and HST-GO-13650.01 to the University of Colorado at Boulder.   Data for the MUSCLES Treasury Survey were also acquired as part of $Chandra$ and $XMM$ guest observing programs, supported by $Chandra$ grants GO4-15014X and GO5-16155X from Smithsonian Astrophysical Observatory and NASA $XMM$ grant NNX16AC09G to the University of Colorado at Boulder.   This work is based in part upon observations obtained with the Apache Point Observatory 3.5-meter and 0.5-meter telescopes, which are owned and operated by the Astrophysical Research Consortium.
KF thanks Evgenya Shkolnik for enjoyable discussions about low-mass stars and Jorge Sanz-Forcada for assistance with the absolute flux levels of the {\tt X-exoplanets}
model spectra.  The MUSCLES team also thanks STScI program coordinator Amber Armstrong for her long hours spent scheduling these complicated coordinated observations.   PCS gratefully acknowledges an ESA Research Fellowship.  SLH acknowledges support from NSF grant AST 13-11678.  FT is supported by the National Natural Science Foundation of China (41175039), the Startup Fund of the Ministry of Education of China (20131029170), and the Tsinghua University Initiative Scientific Research Program.


Author Affiliations: \\
$^{1}$Laboratory for Atmospheric and Space Physics, University of Colorado, 600 UCB, Boulder, CO 80309;  kevin.france@colorado.edu \\
$^{2}$Center for Astrophysics and Space Astronomy, University of Colorado, 389 UCB, Boulder, CO 80309 \\
$^{3}$European Space Research and Technology Centre (ESA/ESTEC), Keplerlaan 1, 2201 AZ Noordwijk, The Netherlands \\
$^{4}$Department of Astronomy, University of Washington, Box 351580, Seattle, WA 98195, USA \\
$^{5}$Department of Astronomy, C1400, University of Texas at Austin, Austin, TX 78712 \\ 
$^{6}$JILA, University of Colorado and NIST, 440 UCB, Boulder, CO 80309 \\ 
$^{7}$Exoplanets and Stellar Astrophysics Laboratory, 
 	NASA Goddard Space Flight Center, Greenbelt, MD 20771 \\ 
$^{8}$Instituto de Astronom\'ia y F\'isica del Espacio (UBA-CONICET) and Departamento de F\'isica (UBA) ,CC.67, suc. 28, 1428, Buenos Aires, Argentina \\ 
$^{9}$Department of Physics \& Astronomy, Western Washington University, Bellingham, WA 98225 \\  
$^{10}$NSF Astronomy and Astrophysics Postdoctoral Fellow \\ 
$^{11}$NorthWest Research Associates, 3380 Mitchell Lane, Boulder, CO 80301-2245 \\ 
$^{12}$Carl Sagan Institute, Cornell University, Ithaca, 14850, NY, USA \\ 
$^{13}$Department of Astronomy, University of Maryland, College Park, MD 20742, USA \\ 
$^{14}$Laboratoire Lagrange, Universite de Nice-Sophia Antipolis, Observatoire de la Cote d'Azur, CNRS, Blvd de l'Observatoire, CS 34229, 06304 Nice cedex 4, France \\
$^{15}$Astronomy Department and Van Vleck Observatory, Wesleyan University, Middletown, CT 06459-0123, USA \\
$^{16}$Department of Earth and Environmental Sciences, Irvine Building, University of St Andrews, St Andrews KY16 9AL, UK \\ 
$^{17}$Ministry of Education Key Laboratory for Earth System Modeling, Center for Earth System Science, Tsinghua University, Beijing 100084, China \\
$^{18}$Instituto de Astronom\'ia y F\'isica del Espacio (UBA-CONICET) and UNTREF,CC.67, suc. 28, 1428, Buenos Aires, Argentina \\ 
$^{19}$The Adler Planetarium, 1300 S Lakeshore Dr, Chicago IL 60605 

\appendix

\section{MUSCLES Targets}

{\it GJ 1214}~--~ GJ 1214 is a late M dwarf (M4.5V) at a distance of 14.6 pc, making it the coolest and most distant M dwarf in the MUSCLES Treasury Survey.  It has an age of~6~$\pm$~3 Gyr~\citep{charbonneau09}, a roughly solar metallicity ([Fe/H]~=~+0.05; Neves et al. 2014), and shows signs of optical flare activity~\citep{kundurthy11}.  Coronal emission from GJ 1214 was recently observed by {\it XMM-Newton}~\citep{lalitha14} with log$_{10}$$L_{X}$~$=$~25.87 erg s$^{-1}$.  GJ 1214b is a transiting super-Earth ($M_{plan}$~$\approx$~6.5~$M_{\oplus}$, $a_{plan}$~=~0.014 AU; Charbonneau et al. 2009), possibly harboring a dense, water-rich~\citep{bean10,desert11} and likely cloudy~\citep{kreidberg14} atmosphere.

{\it GJ 876}~--~ GJ 876 is an M5 dwarf at a distance of 4.7 pc.  GJ 876 has super-solar metallicity ([Fe/H] = 0.14; Neves et al. 2014); differing estimates on the stellar rotation period (40~$\leq$~$P_{*}$~$\lesssim$~97 days) result in large uncertainties in the age estimate for this system, 0.1~--~5 Gyr~\citep{rivera05,rivera10,correia10}.\nocite{johnson09,rojas10}  While the star would be characterized as weakly active based on its H$\alpha$ absorption spectrum, UV and X-ray observations show the presence of an active upper atmosphere (Walkowicz et al. 2008; France et al. 2012a; log$_{10}$$L_{X}$~$=$~26.48 erg s$^{-1}$; Poppenhaeger et al. 2010).  GJ 876 has a rich planetary system with four planets ranging from a super-Earth (GJ 876d, $M_{plan}$~$\approx$~6.6~$M_{\oplus}$) in a short-period orbit ($a_{plan}$~=~0.02 AU; Rivera et al. 2010) to two Jovian-mass planets in the HZ (GJ 876b, $M_{plan}$~$\approx$~2.27~$M_{Jup}$, $a_{plan}$~=~0.21 AU; GJ 876c, $M_{plan}$~$\approx$~0.72~$M_{Jup}$, $a_{plan}$~=~0.13 AU; Rivera et al. 2010).

{\it GJ 581}~--~ GJ 581 is an M5 dwarf at a distance of 6.3 pc.  It is estimated to have an age of 8~$\pm$~1 Gyr~\citep{selsis07} and a somewhat subsolar metallicity, [Fe/H]~=~$-$0.20~\citep{neves14}.  GJ 581 was not detected in early X-ray surveys (log$_{10}$$L_{X}$~$<$~26.89 erg s$^{-1}$; Poppenhaeger et al. 2010), however X-rays were subsequently detected, log$_{10}$$L_{X}$~$=$~26.11 erg s$^{-1}$, by $Swift$~\citep{vitale13}. Its optical spectrum displays H$\alpha$ in absorption, therefore chromospheric and coronal activity are thought to be low for this target.\nocite{poppenhaeger10}  GJ 581 has one of the richest known planetary systems, with possibly up to six planets (four confirmed) including several with Earth/super-Earth masses~\citep{mayor09,tuomi11}.  GJ 581d is a super-Earth ($M_{plan}$~$\approx$~6~$M_{\oplus}$) that resides on the outer edge of the habitable zone (HZ; $a_{plan}$~=~0.22 AU; Wordsworth et al. 2011; von Braun et al. 2011). \nocite{wordsworth11,vonbraun11}  

{\it GJ 176}~--~ GJ 176 is an M2.5 dwarf at a distance of 9.4 pc.  It is estimated to have an age between 0.56 - 3.62 Gyr~\citep{kiraga07,forcada11} and a solar metallicity, [Fe/H]~=~$-$0.01~\citep{neves14}.  GJ 176 has an archival X-ray luminosity of (log$_{10}$$L_{X}$~$=$~27.48 erg s$^{-1}$; Poppenhaeger et al. 2010) and a 39 day rotational period~\citep{robertson15}.\nocite{poppenhaeger10}  GJ 176 b was initially detected as a 24 $M_{\oplus}$ planet in a 10 day orbit~\citep{endl08}, but subsequent analysis of the stellar light-curve has refined this estimate to a super-Earth mass planet (8.3 $M_{\oplus}$) in an 6.6 day orbit~\citep{butler09,forveille09}.   

{\it GJ 436}~--~ GJ 436 is an M3.5 dwarf star located at a distance of 10.3 pc.  It has a 45 day rotation period, a relatively old age ($\sim$~6$^{+4}_{-5}$~Gyr; Torres 2007), and may have a super-solar metallicity ([Fe/H]~=~0.00~--~0.25; Neves et al. 2014; Johnson \& Apps 2009; Rojas-Ayala et al. 2010).  GJ 436 does show signs of an active corona with log$_{10}$$L_{X}$~$=$~27.16 erg s$^{-1}$ (Poppenhaeger et al. 2010), and its chromospheric Ly$\alpha$ emission was previously observed by~\citet{ehrenreich11} and~\citet{kulow14}.   GJ 436 is notable for its well-studied transiting Neptune mass planet~\citep{butler04,pont09}, orbiting at a semi-major axis of ~$\approx$~0.03 AU, interior to its HZ (0.16~--~0.31~AU; von Braun et al. 2012).\nocite{vonbraun12,torres07}  The impact of Ly$\alpha$ on the atmosphere of GJ 436b has been studied by~\citet{miguel15}.  Additional low-mass planets may also be present in this system~\citep{stevenson12}.  

{\it GJ 667C}~--~ GJ 667C (M1.5V) is a member of a triple star system (GJ 667AB is a K3V + K5V binary) at a distance of 6.9 pc.  This 2~--~10 Gyr M dwarf~\citep{anglada_escude12} is metal-poor ([Fe/H] = $-$0.50~Neves et al. 2014, and a similar value of $-$0.59~$\pm$~0.10 based on an analysis of GJ 667AB; Perrin et al. 1988).  GJ 667C may host as many as three planets, including a super-Earth mass planet (GJ 667Cc, $M_{plan}$~$\approx$~4.5~$M_{\oplus}$, $a_{plan}$~=~0.12 AU) orbiting in the HZ (0.11~--~0.23 AU; Anglada-Escud{\'e} et al. 2012).\nocite{perrin88}

{\it GJ 832}~--~ GJ 832 is an M1.5 dwarf at $d$~=~4.9 pc.  GJ 832 is not as well characterized as other targets in our sample; an age determination for this star is not available.  Coronal X-rays have been detected from GJ 832 with log$_{10}$$L_{X}$~$=$~26.77 erg s$^{-1}$ (Poppenhaeger et al. 2010).  This subsolar metallicity star ([Fe/H] = $-$0.17; Neves et al. 2014) hosts two known exoplanets: ``b'', a 0.7~$M_{Jup}$ Jovian planet at $a_{b}$~=~3.6~AU (Bailey et al. 2009) and ``c'', a 5.4 $M_{\oplus}$ super-Earth planet in the Habitable Zone~($a_{c}$~=~0.16~AU; Wittenmyer et al. 2014).\nocite{bailey09,wittenmyer14}

{\it HD 85512}~--~ HD 85512 is a K6 dwarf at a distance of 11.2 pc.  It is estimated to have an age of 5.6~$\pm$~0.1 Gyr and a subsolar metallicity, [Fe/H]~=~$-$0.26~\citep{tsantaki13}, based on its 47 day rotation period. There is no previously published X-ray detection of HD 85512, the {\it XMM-Newton} luminosity is log$_{10}$$L_{X}$~$\sim$~26.5 erg s$^{-1}$, (Loyd et al. 2016, Brown et al. 2016).   HD 85512 b is a super-Earth mass planet (3.6 $M_{\oplus}$) orbiting with a semi-major axis of 0.26 AU~\citep{pepe11}.  

{\it HD 40307}~--~ HD 40307 is a K2.5 dwarf at a distance of 12.9 pc.  It is estimated to have an age of 4.5 Gyr~\citep{barnes07} and a subsolar metallicity, [Fe/H]~=~$-$0.36~\citep{tsantaki13}.  HD 40307 has a 48 day rotation period~\citep{mayor09} and an archival X-ray luminosity of log$_{10}$$L_{X}$~$=$~26.99 erg s$^{-1}$ (Poppenhaeger et al. 2010).   HD 40307 hosts somewhere between 3 and 6 planets, including several of super-Earth mass (Table 1; Mayor et al. 2009, Tuomi et al. 2013).\nocite{tuomi13,mayor09}   

{\it $\epsilon$~Eri}~--~ $\epsilon$~Eri is a K2 dwarf at a distance of 3.2 pc.  It is one of the best-studied active K stars~\citep{dring97, ness02,jeffers14}.    It is a relatively young star with an age $\approx$~0.44 Gyr~\citep{barnes07}, an 11.68 day rotation period~\citep{rueedi97}, and a slightly subsolar metallicity, [Fe/H]~=~$-$0.15~\citep{tsantaki13}.  $\epsilon$~Eri displays magnetic cycles that suggest it may have been in a relatively ``inactive'' state during the MUSCLES observations~\citep{metcalfe13}.  Archival spectra from {\it XMM-Newton} show an X-ray luminosity in the 0.2~--~2.0 keV energy band of log$_{10}$$L_{X}$~$=$~28.22 erg s$^{-1}$ (Poppenhaeger et al. 2010) that grows to  log$_{10}$$L_{X}$~$=$~29.32 erg s$^{-1}$ when the full range covered by the $Chandra$ LETGS is included (0.07~--~2.5 keV; Ness et al. 2002).   $\epsilon$~Eri hosts a $\sim$~1.1 $M_{Jup}$ planet in a 3.4 AU semi-major axis orbit~\citep{hatzes00,butler06}.  

{\it HD 97658}~--~ HD 97658 is a K1 dwarf at a distance of 21.1 pc.  It is estimated to have an age of 3.8~$\pm$~2.6 Gyr~\citep{bonfanti15} and a subsolar metallicity [Fe/H]~=~$-$0.26 \citep{valenti05}.  There are no published X-ray data of HD 97658 in the literature, but X-ray observations are scheduled as part of $Chandra$'s Cycle 16. HD97658 hosts a super-Earth mass planet (7.9 $M_{\oplus}$) in a short-period orbit ($a$~=~0.08~AU; Dragomir et al. 2013).




 \scriptsize
\hspace{-0.0in}
\begin{table*}[t!]\vspace{-0.00in}\footnotesize
\centering
\caption{MUSCLES Treasury Survey~--~Target List
}\vspace{-0.0in}
\begin{tabular}{l@{\hspace{-1ex}}r@{\hspace*{3ex}}ccccccc}
\hline \hline
Star 	& \multicolumn{1}{c}{Distance} 	& Type &	   T$_{eff}$                &    P$_{rot}$     &    $\langle$$r_{HZ}$$\rangle$$^{a}$             &  \multicolumn{1}{c}{Exoplanet Mass} 	&	 Semi-major Axis & Ref.$^{b}$	 \\
	&  \multicolumn{1}{c}{(pc)}		     &	               &  (K)                        &  (days)             &     (AU)  &  \multicolumn{1}{c}{$M$ sin $i$ (M$_{\oplus}$)}               & (AU)	&                               	\\
\hline
GJ 1214	& 14.6 	& M4.5			       & 2935  &  53  &  0.096	& {\bf 6.4} 				&	0.0143		& 1,2,3		  \\	
GJ 876	& 4.7		& M5			& 3062 &  96.7  &  0.178 & 615, 194,		 	&	0.208, 0.130, 		& 3,4,5 \\
		&  		&  			&    &     &   	& {\bf 5.7}, 12.4 			&	0.021, 0.334			&    \\
GJ 581	& 6.3		& M5			& 3295  &  94.2  & 0.146   & 15.9, {\bf 5.4},		 	&	0.041, 0.073, 	& 3,4,6	  \\
		&  		&  			&    &    &   	& {\bf 6.0}, {\bf 1.9 }			&	{\bf 0.218}, 0.029		& 	 \\
GJ 436	& 10.3	& M3.5 		             & 3281  &  48  &  0.211 	& 23 				&	0.0287				& 3,4,7		 \\
GJ 176	& 9.4		& M2.5 	        	& 3416  &  38.9  &  0.262	& {\bf 8.3} 				&	0.066			& 3,4,8		 \\
GJ 667C	& 6.9		& M1.5 		      & 3327  &  105  &  0.172 	& {\bf 5.7}, {\bf 4.4}			 		&	0.049,{\bf 0.123}	& 3,1,9	  \\
GJ 832	& 4.9 	& M1.5	 		      & 3816  &  40$^{c}$ &  0.235  & 203, 	{\bf 5.4}			&	3.6, {\bf 0.16}			& 3,4		  \\
HD 85512	& 11.2 	& K6			& 4305  &  47.1  &  0.524 	& {\bf 3.5} 				&	0.26		& 3,10,11			 \\

HD 40307		& 12.9 	& K2.5		& 4783 &  48 &  0.674 	& {\bf 4.1}, {\bf 6.7},  				&	0.047, 0.080,		& 3,10,12	\\
		&  		&  			&   &    &   	& 9.5, {\bf 3.5}, 			&	0.132, 0.189, 			& 		 \\
		&  		&  			&    &     &   	& {\bf 5.1}, {\bf 7.0 }			&	0.247, 0.600			& 		 \\
$\epsilon$ Eri	& 3.2	 	& K2 	& 5162  &  11.7  &  0.726	& $\sim$400 				&	3.4			& 3,10,13	   \\
HD 97658	& 21.1 	& K1			& 5156  &  38.5  &  0.703	& {\bf 6.4} 				&	0.080			& 3,14,15		  \\
\hline
 \end{tabular}
\\ $^{a}$~--~Habitable zone distance is defined as the average of the runaway greenhouse and maximum greenhouse limits (Kopparapu et al. 2014).\\ 
$^{b}$~--~References: 1.~\citet{neves14}, 2.~\citet{berta11}, 3. Simbad parallax distance, 4.~\citet{gaidos14}, 5.~\citet{rivera10}, 6.~\citet{robertson14}, 7.~\citet{demory07},  8.~\citet{kiraga07},9.~\citet{anglada13},10.~\citet{tsantaki13},11.~\citet{pepe11},12.~\citet{mayor09},
13.~\citet{donahue96},14.~\citet{valenti05},15.~\citet{henry11} \\
$^{c}$~--~GJ 832 has no published rotation period, we assume a relatively short period due to the persistence of UV flare activity in this star.  \\
\end{table*}

\begin{table}
\centering
\caption{Emperical Log-log Relations,  log$_{10}$$y$~=~($m$~$\times$~log$_{10}$$x$)~+~$b$    \label{lya_lines}}
 \begin{tabular}{lc|cc} 
\hline \hline
$y$   & $x$ & $m$   &  $b$  \\
\hline 
$L$(FUV)$^{a}$     &   $L$(\ion{C}{4})     &   1.12~$\pm$~0.27    &    $-$1.01~$\pm$~6.86      \\
$L$(FUV)     &   $L$(\ion{Mg}{2})  &   0.75~$\pm$~0.06    &     7.56~$\pm$~1.52     \\
$L$(XUV)     &   $L$(\ion{C}{4})     &   0.97~$\pm$~0.19    &    2.72~$\pm$~4.91      \\
$L$(XUV)     &   $L$(\ion{Mg}{2})  &  0.58~$\pm$~0.08    &    11.98~$\pm$~2.12      \\
\hline
FUV/NUV     &   $\langle$$r_{HZ}$$\rangle$     &   $-$2.42~$\pm$~0.30    &    $-$2.31~$\pm$~0.18    \\
HZ FUV Flux$^{a}$     &   $\langle$$r_{HZ}$$\rangle$     &   0.51~$\pm$~0.16    &   1.71~$\pm$~0.10    \\
HZ XUV Flux$^{a}$   &   $\langle$$r_{HZ}$$\rangle$     &   $-$0.01~$\pm$~0.22   &  1.41~$\pm$~0.14     \\
\hline
$L$(\ion{N}{5})/$L_{Bol}$  &   $M_{plan}$/$a_{plan}$     &   0.64~$\pm$~0.24    &    $-$8.53~$\pm$~0.56   \\
\hline
 \end{tabular}
\\ $^{a}$~--~All luminosities in units of (erg s$^{-1}$).   \\
$^{b}$~--~Habitable Zone fluxes in units of (erg cm$^{-2}$ s$^{-1}$), $\langle$$r_{HZ}$$\rangle$ and $a_{plan}$ in units of (AU), $M_{plan}$ in units of (M$_{\oplus}$).    
\end{table}

\begin{table} 
\tabletypesize{\normalsize}
\centering
\caption{MUSCLES Broadband and Emission Line Luminosity$^{a,b,c}$. \label{lya_lines}}
 \begin{tabular}{l|cccccccc}
Target &  $L_{Bol}$ & $f$(XUV) &  $f$(FUV) &  $f$(NUV)    &
     $f$(Ly$\alpha$) &  $f$(\ion{N}{5}) &  $f$(\ion{C}{4}) &  $f$(\ion{Mg}{2})  \\
\hline
GJ1214  &  31.15  &  -4.50  &  -4.45  &  -4.20  &  -4.52  &  -6.48  &  -6.03  &  -5.53  \\  
GJ176  &  32.12  &  -4.46  &  -4.40  &  -3.81  &  -4.51  &  -6.61  &  -6.01  &  -4.91  \\  
GJ436  &  32.02  &  -4.41  &  -4.52  &  -4.02  &  -4.59  &  -6.93  &  -6.54  &  -5.33  \\  
GJ581  &  31.64  &  -4.50  &  -4.84  &  -4.18  &  -4.90  &  -7.23  &  -6.70  &  -5.69  \\  
GJ667C  &  31.70  &  -4.21  &  -4.17  &  -3.86  &  -4.23  &  -7.11  &  -6.47  &  -5.33  \\  
GJ832  &  31.79  &  -4.30  &  -4.29  &  -3.60  &  -4.36  &  -6.80  &  -6.46  &  -5.07  \\  
GJ876  &  31.69  &  -4.04  &  -4.53  &  -4.33  &  -4.68  &  -6.24  &  -5.91  &  -5.77  \\  
HD40307  &  32.98  &  -4.57  &  -4.31  &  -2.41  &  -4.37  &  -8.20  &  -7.21  &  -5.15  \\  
HD85512  &  32.78  &  -4.82  &  -4.45  &  -3.01  &  -4.51  &  -7.79  &  -6.89  &  -4.93  \\  
HD97658  &  33.12  &  -4.61  &  -4.36  &  -2.05  &  -4.43  &  -7.86  &  -6.97  &  -5.02  \\  
$\epsilon$~Eri  &  33.08  &  -4.15  &  -4.10  &  -1.97  &  -4.20  &  -6.97  &  -6.13  &  -4.30  \\  
\hline
\end{tabular}
\\ $^{a}$~--~Flux measurements are averaged over all exposure times for individual observations, and broadband SEDs are constructed as described in~\citet{loyd15}. 
\\ $^{b}$~--~All quantities presented as log$_{10}$($L_{Bol}$)  or  $f$(band) = log$_{10}$($L$(band)/$L_{Bol}$).   
\\ $^{c}$~--~Broadband bandpasses are defined as: Bol $\Delta$$\lambda$ = 5\AA~--~$\infty$, XUV $\Delta$$\lambda$ = 5~--~911\AA, FUV $\Delta$$\lambda$ = 912~--~1700\AA\ (including Ly$\alpha$), NUV $\Delta$$\lambda$ = 1700~--~3200\AA. 
\end{table}

\begin{table}   
\tabletypesize{\normalsize}
\caption{Comparison of MUSCLES \ion{C}{3} $\lambda$1175 emission and {\tt X-exoplanets} spectral synthesis prediction.}
\centering
\begin{tabular}{lcc}
  Star & F(\ion{C}{3}), MUSCLES & F(\ion{C}{3}), {\tt X-exoplanets}  \\
\hline
GJ 876     &   6.5~$\times$~10$^{-4}$    &   1.9~$\times$~10$^{-5}$       \\
GJ 436     &   4.2~$\times$~10$^{-5}$ &   8.0~$\times$~10$^{-6}$         \\
$\epsilon$ Eri     &  1.8~$\times$~10$^{-2}$   &   6.6~$\times$~10$^{-4}$       \\
\hline
\end{tabular}
\\$^{a}$~--~All fluxes in units of (photons  s$^{-1}$ cm$^{-2}$).   
\end{table}


\bibliography{ms_M22_emapj}

\end{document}